\documentclass[10pt,leqno]{amsart}

\usepackage[utf8]{inputenc}
\usepackage[T1]{fontenc}
\usepackage{graphicx}
\usepackage{indentfirst,csquotes}

\usepackage{amssymb,amsthm,amsmath}
\usepackage{bm}
\usepackage{booktabs}
\usepackage{multirow}
\usepackage{dcolumn}
\usepackage{algpseudocode}
\usepackage[numbers,sort&compress]{natbib}
\usepackage{xcolor}
\usepackage{paralist}
\usepackage{titlesec}
\usepackage{fancyhdr}
\usepackage{etoolbox}
\usepackage{threeparttable}
\usepackage{hyperref}

\makeatletter
\providecommand{\@secnumpunct}{.}
\makeatother

\hypersetup{
  colorlinks=true,
  linkcolor=black,
  citecolor=black,
  filecolor=black,
  urlcolor=black
}

\titleformat{\section}[display]
  {\normalfont\huge\bfseries\centering}
  {\thesection}{10pt}{\Large}
\titlespacing*{\section}{0pt}{0ex}{0ex}

% ...
\newcounter{algorithm}
\renewcommand{\thealgorithm}{\arabic{algorithm}}

\newenvironment{algorithmblock}[2][]{%
  \refstepcounter{algorithm}%
  \par\medskip
  \noindent\textbf{Algorithm~\thealgorithm. #2}%
  \if\relax\detokenize{#1}\relax\else\label{#1}\fi
  \par\smallskip
  \hrule
  \smallskip
  \begingroup
  \small
}{%
  \par
  \endgroup
  \smallskip
  \hrule
  \medskip
}

\begin{document}
\title{Nyquist-Sampled Time-Domain Adjoint FDTD for Memory-Efficient Broadband Nanophotonic Inverse Design}

\author[M. Park]{Mingyu Park}
\address{Department of Electronic Engineering, Hanyang University, Seoul 04763, Republic of Korea}

\author[O. D. Miller]{Owen D. Miller$^\ast$}
\address{Department of Applied Physics and Energy Sciences Institute, Yale University, New Haven, CT 06520, USA}
% \email{owen.miller@yale.edu}

\author[H. Chung]{Haejun Chung$^\ast$}
\address{Department of Electronic Engineering, Hanyang University, Seoul 04763, Republic of Korea}
\address{Department of Artificial Intelligence Semiconductor Engineering, Hanyang University, Seoul 04763, Republic of Korea}
\address{Department of Artificial Intelligence, Hanyang University, Seoul 04763, Republic of Korea}
% \email{haejun@hanyang.ac.kr}

\date{\today}
\maketitle

\let\thefootnote\relax
\footnotetext{$^\ast$Corresponding authors. Keywords: Time-domain adjoint method, Inverse design, Topology optimization, Electromagnetic design, FDTD.}

\begin{abstract}
Adjoint optimization is a cornerstone of broadband nanophotonic inverse design, but conventional time-domain implementations face a severe memory bottleneck because they retain forward-field histories at every finite-difference time-domain (FDTD) time step. Here, we show that this full time-step storage is unnecessary for broadband design objectives because the underlying fields are band-limited. By storing forward fields only at Nyquist-compliant temporal intervals and using the resulting sparse field history during the reverse-time adjoint pass, the proposed method enables on-the-fly gradient accumulation without retaining full forward- or adjoint-field histories. This Nyquist-sampled adjoint FDTD framework preserves the two-simulation scaling of time-domain adjoint optimization while substantially reducing the dominant field-storage cost. Because the broadband gradient is evaluated directly in the time domain, with no spectral discretization, the per-iteration cost is also independent of the number of target frequencies---in contrast to frequency-sampled adjoint formulations, whose cost grows with spectral sampling density. Gradient verification confirms that Nyquist-compliant sampling reproduces conventional full-storage adjoint gradients with negligible error, whereas undersampling beyond the Nyquist limit produces aliasing-induced gradient degradation. Across four two-dimensional broadband nanophotonic benchmarks and a fully three-dimensional metalens, the method maintains gradient fidelity and optimized device performance while reducing dominant field-storage memory by more than $100\times$ relative to full-history storage in a prototypical example. These results suggest that the principal memory barrier in broadband time-domain adjoint FDTD is not an intrinsic requirement of gradient evaluation but rather a consequence of redundant temporal field storage, thereby opening a practical route to large-scale three-dimensional nanophotonic inverse design.
\end{abstract}

\bigskip
\maketitle
\section{Introduction}

Adjoint optimization has emerged as a powerful framework for large-scale nanophotonic inverse design, enabling the efficient optimization of structures with many degrees of freedom. However, broadband or many-wavelength design remains computationally challenging. Frequency-domain adjoint methods resolve the gradient at discretely sampled frequencies, requiring either a separate adjoint simulation at each frequency or a single adjoint simulation whose synthesized broadband source---and with it the simulation window---lengthens in proportion to the sampling density; either way, they incur computation times that scale approximately linearly with the number of frequency samples. Time-domain adjoint methods can overcome this runtime scaling by evaluating the broadband gradient directly in the time domain, with no spectral discretization, with only one forward and one adjoint simulation; the price is that conventional implementations require storing the full temporal field history throughout the design region, resulting in prohibitively large memory consumption for large-scale, three-dimensional problems. Here, we propose a fully time-domain adjoint optimization framework that applies a Nyquist--Shannon downsampling scheme to the stored forward fields. By retaining only Nyquist-satisfying forward-field samples and accumulating adjoint-gradient contributions on the fly, the proposed method preserves broadband gradient accuracy while substantially reducing memory requirements.
This approach retains the computational efficiency of time-domain adjoint optimization: its per-iteration cost is independent of the number of sampled frequencies, an advantage over frequency-sampled adjoint approaches that exceeds an order of magnitude at the densest spectral samplings considered here and grows in proportion to the sampling density.
Concurrently, it lowers memory consumption by up to $107\times$ compared with conventional full-history time-domain adjoint implementations, reducing the stored forward-field data to the information-theoretic floor set by the field bandwidth.

Adjoint-gradient-based design methods have been extensively studied across multiple disciplines, including mechanical engineering~\cite{zhou2008design}, aerospace engineering~\cite{jameson1988aerodynamic,aage2017giga}, computational electromagnetics~\cite{taflove2005computational}, and photonics~\cite{burger2004inverse}. These methods are closely related to error backpropagation originating in the 1980s~\cite{werbos1994roots,rumelhart1986learning}, but in physics-based design, they are more commonly formulated within the framework of adjoint sensitivity analysis for PDE-constrained optimization. Depending on the design parameterization and update strategy, related approaches are often referred to as topology optimization~\cite{bendsoe1999material,christiansen2021inverse}, shape optimization~\cite{lalau2013adjoint}, or level-set
optimization~\cite{otomori2012topology}. In nanophotonics, adjoint-based inverse design has become an important computational design paradigm because it enables efficient sensitivity evaluation for a large number of design degrees of freedom. Early developments drew on topology optimization and electromagnetic adjoint formulations, and subsequent studies have demonstrated high-performance photonic devices, including wavelength demultiplexers~\cite{piggott2015inverse}, metagratings~\cite{burger2004inverse}, metalenses~\cite{Chung:20}, solar cells~\cite{ganapati2013light}, and integrated photonic components~\cite{nikkhah2024inverse}. These works have frequently achieved state-of-the-art performance within their respective device classes, establishing adjoint optimization as an efficient and scalable approach for photonic design. More recently, automatic-differentiation-based frameworks have further lowered the barrier to adjoint optimization by enabling gradients and adjoint sources to be generated directly from user-defined figures of merit, reducing the need for manual derivations and facilitating broader adoption across photonics and optics.

The computational speed of adjoint-based photonic design has benefited substantially from advances in high-performance computing hardware, particularly from improvements in memory bandwidth. Full-wave solvers for Maxwell’s equations, such as FDTD, are often memory-bandwidth limited because field components must be repeatedly read from and written to memory at every time step. Conventional DDR memory provides bandwidth on the order of tens of GB/s per module, whereas modern high-bandwidth-memory architectures integrated with GPUs can deliver several TB/s of bandwidth, enabling substantial acceleration of both forward and adjoint electromagnetic simulations. Nevertheless, relying solely on hardware scaling is unlikely to yield a sustainable solution for large-scale photonic inverse design, as the per-pin data rate of the high-bandwidth memory starts to saturate. This limitation is particularly important for emerging applications such as large-area metasurfaces, photonic integrated circuits, and optical computing platforms, where device footprints can span hundreds to many thousands of wavelengths and require extremely large computational domains. These considerations motivate the development of algorithmic strategies that are not only faster but also more memory efficient. In this context, reducing redundant field storage and minimizing data movement are essential to extending adjoint optimization to practical, large-scale, three-dimensional photonic systems.

In this work, we propose a fully time-domain adjoint optimization framework combined with a Nyquist-theorem-based downsampling strategy for memory-efficient broadband photonic inverse design. The key observation is that the electromagnetic field histories generated by band-limited excitations need not be stored at every FDTD time step to recover the frequency-domain information required for adjoint-gradient evaluation. According to the Nyquist–Shannon sampling theorem, these field histories can be represented without loss of information when sampled at a rate at least twice the maximum relevant frequency. By exploiting this principle, the proposed method stores forward-field data only at Nyquist-satisfying time intervals and accumulates gradient contributions on the fly during the adjoint simulation, evaluating the broadband gradient---an overlap integral over the full target band---exactly by quadrature on the downsampled grid. As a result, it preserves the broadband efficiency of time-domain adjoint optimization while reducing redundant temporal field storage by up to $107\times$ relative to full-history storage. 

The remainder of this paper is organized as follows. Section~\ref{sec:method} presents the FDTD and topology-optimization foundations, reviewing Maxwell’s equations, their FDTD discretization, and the design parameterization. Section~\ref{sec:adjoint-review} compares the frequency- and time-domain adjoint formulations, organized by where each pays the cost of dense spectral sampling. Section~\ref{sec:proposed} derives the Nyquist-based downsampling strategy and introduces an on-the-fly gradient accumulation scheme, analyzing their effects on memory scaling and gradient accuracy. Section~\ref{sec:results} validates the proposed framework through numerical experiments on representative nanophotonic devices, including gradient verification against the conventional full-storage method, memory-scaling analysis as a function of the downsampling factor, and a fully three-dimensional broadband metalens optimization. Finally, Section~\ref{sec:conclusion} summarizes the computational advantages of the proposed approach and discusses its potential extension to other time-domain adjoint problems beyond electromagnetics.

\begin{figure}[!htbp]
\centering
\includegraphics[width=\textwidth]{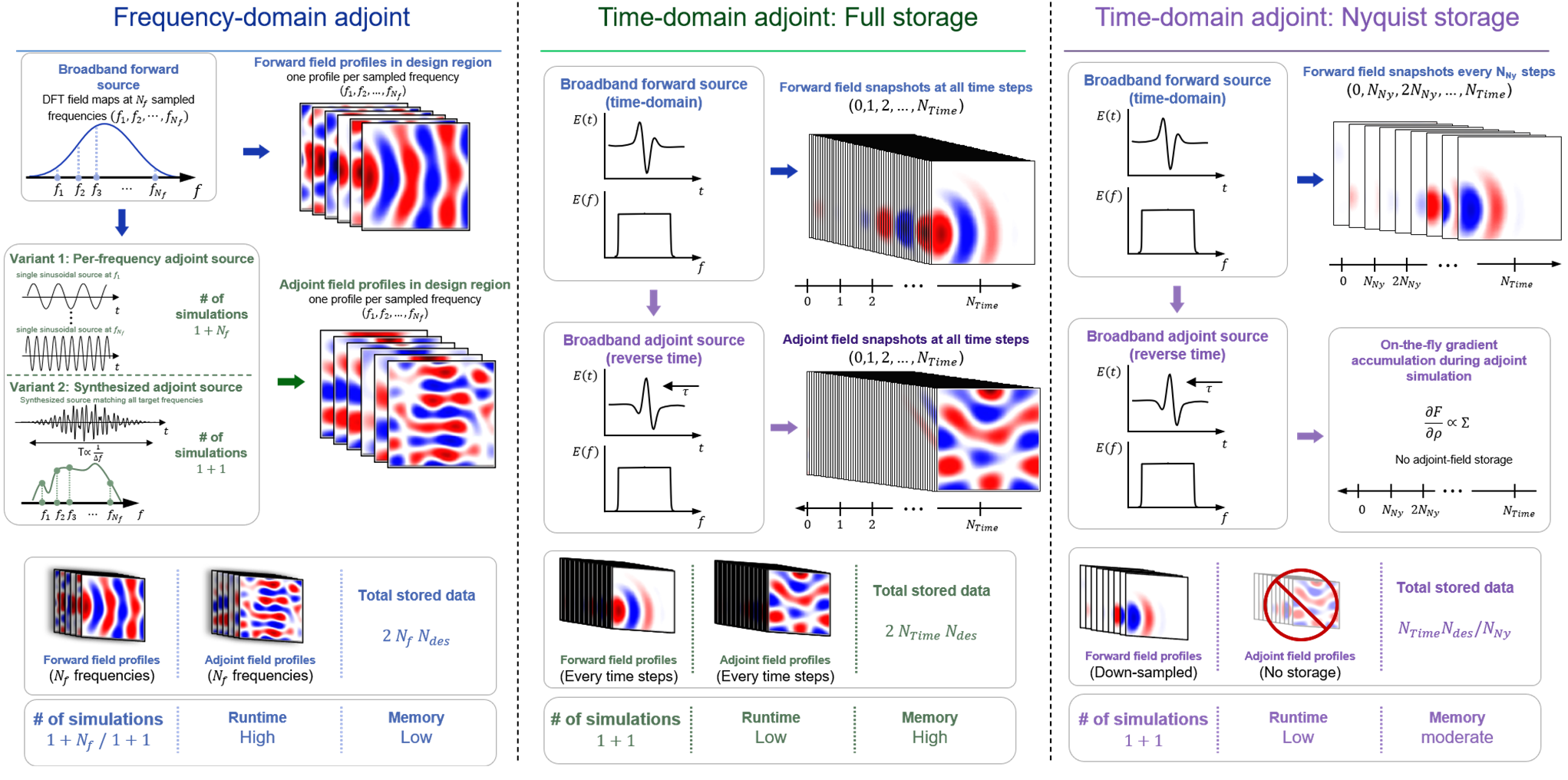}
\caption{
Comparison of frequency-domain and time-domain adjoint formulations for broadband nanophotonic inverse design.
(a) The frequency-domain adjoint formulation stores forward and adjoint field profiles at $N_f$ sampled frequencies, with a field-storage cost proportional to $2N_fN_{\mathrm{des}}$. Two variants are illustrated: a per-frequency approach requiring $1+N_f$ simulations and a synthesized-source approach retaining a $1+1$ simulation structure by combining the frequency-dependent adjoint excitations into a single broadband source, whose duration increases as the spectral spacing $\Delta f$ is refined.
(b) The conventional time-domain adjoint formulation uses one broadband forward simulation and one time-reversed adjoint simulation, independent of $N_f$, but stores forward and adjoint field histories at every FDTD time step, resulting in a storage cost proportional to $2N_{\mathrm{des}}N_{\mathrm{time}}$ for the full-storage implementation shown.
(c) The proposed Nyquist-sampled time-domain adjoint formulation retains the $1+1$ simulation structure while storing forward-field snapshots only every $N_{\mathrm{Ny}}$ FDTD steps and accumulating the gradient on the fly during the adjoint simulation. This eliminates adjoint-field storage and reduces the dominant storage requirement to approximately $N_{\mathrm{des}}N_{\mathrm{time}}/N_{\mathrm{Ny}}$.}

\label{fig:overview}
\end{figure}

\section{Adjoint-based topology optimization in FDTD}
\label{sec:method}
% ============================================================

This section establishes the numerical and design foundations of the proposed
Nyquist-sampled time-domain adjoint method. We first introduce the FDTD
discretization and the density-based topology optimization formulation,
which provide the numerical and design foundations of this work. 
The frequency- and time-domain adjoint formulations, together with their
runtime and memory trade-offs, are compared in Section~\ref{sec:adjoint-review}.
The proposed method, developed in
Section~\ref{sec:proposed}, reduces the memory burden of time-domain adjoint
optimization by storing forward fields at Nyquist-sampled temporal intervals
rather than at every FDTD time step and accumulating the design gradient on
the fly during the adjoint simulation~\cite{ayachit2015paraview,bauer2016situ}.

%================================
\subsection{FDTD formulation and temporal oversampling}
\label{subsec:fdtd}
%================================

The proposed Nyquist-sampled field-storage scheme is not restricted to FDTD
and can be incorporated into other time-marching electromagnetic solvers that
provide access to the forward fields during propagation, including the
time-domain finite-element method, the discontinuous Galerkin time-domain
method, and the finite-volume time-domain method. In this work, we use the FDTD method~\cite{yee1966numerical,taflove2005computational}
as a representative implementation. FDTD advances Maxwell's curl equations,
\begin{equation}
  \frac{\partial \mathbf{B}}{\partial t} = -\nabla\times\mathbf{E},
  \qquad
  \frac{\partial \mathbf{D}}{\partial t} = \nabla\times\mathbf{H}
    - \mathbf{J}_s,
  \label{eq:maxwell}
\end{equation}
on a staggered Yee grid using a leapfrog time-marching scheme~\cite{yee1966numerical}.

The time step $\Delta t$ must satisfy the Courant--Friedrichs--Lewy (CFL)
stability condition~\cite{de2013courant}. In practical FDTD simulations,
the spatial grid resolution is also chosen to satisfy the points-per-wavelength accuracy requirement~\cite{taflove2005computational}, which further
constrains the allowable time step through the CFL condition. As a result,
FDTD often samples the electromagnetic fields at a temporal rate much higher
than that required to represent the bandwidth-limited source and response.
This oversampling creates substantial redundancy in the stored forward-field
history and constitutes the main source of memory inefficiency addressed in
this work.

We truncate the computational domain using perfectly matched layers~\cite{berenger1994perfectly,sullivan1996simplified} and apply periodic
boundary conditions where appropriate. Standard references provide further
details on the FDTD update equations, stability criteria, and boundary
implementations~\cite{taflove2005computational,gedney2011introduction}.

% %================================
\subsection{Design parameterization and binary projection}
\label{subsec:topopt}
%================================

We use density-based topology optimization~\cite{bendsoe1999material} to
parameterize the inverse design problem. A continuous density field
$\rho(\mathbf{r})\in[0,1]$ defines the material distribution in the design
region $\Omega_D$, where the limiting values $\rho=0$ and $\rho=1$
correspond to the two constituent materials.

To obtain fabrication-compatible binary designs, we transform the density
field using a smooth Heaviside projection~\cite{neves2002topology}. The
projected density $\hat{\rho}$ determines the local relative permittivity as
\begin{equation}
  \varepsilon_r(\mathbf{r}) =
  \varepsilon_{\min}
  + \hat{\rho}(\mathbf{r})
  \left(\varepsilon_{\max}-\varepsilon_{\min}\right).
  \label{eq:eps_interp}
\end{equation}
This projection allows the optimization to retain differentiability while
progressively favoring near-binary material layouts.

We obtain the gradient with respect to the design density by applying the chain rule through the permittivity interpolation and projection steps. The Nyquist-sampled adjoint method provides the electromagnetic sensitivity with respect to the material distribution, which we then backpropagate to $\rho(\mathbf{r})$. While we adopt density-based topology optimization in this study, the proposed adjoint framework is compatible with any differentiable parameterization that links design variables to material properties~\cite{molesky2018inverse,christiansen2021inverse}.

%================================
\subsection{Adjoint optimization problem formulation}
\label{subsec:opt-problem}
%================================

We formulate the inverse design problem as the maximization of an objective functional $F[\mathbf{E},\mathbf{H}]$ subject to the time-domain Maxwell equations:
\begin{equation}
  \begin{aligned}
    \max_{\rho}\quad
    & F[\mathbf{E},\mathbf{H}] \\
    \text{subject to}\quad
    & \text{Maxwell's equations and boundary conditions in } \Omega, \\
    & \mathbf{D}=\varepsilon(\hat{\rho})\mathbf{E},
      \qquad \mathbf{B}=\mu\mathbf{H}, \\
    & 0\le\rho(\mathbf{r})\le 1,
      \qquad \mathbf{r}\in\Omega_D .
  \end{aligned}
  \label{eq:opt-cont}
\end{equation}

The design density $\rho(\mathbf{r})$ defines the material distribution in $\Omega_D$ through the projected density $\hat{\rho}$ and the corresponding permittivity $\varepsilon(\hat{\rho})$. In Section~\ref{sec:results}, we define $F$ in the time domain so that the optimization directly targets broadband focusing or transmission performance using field samples measured at specified monitor locations or regions~\cite{Chung:20,park2026multi}.

\section{Adjoint formulations and memory-scaling limitations}
\label{sec:adjoint-review}

In this section, we compare adjoint formulations for broadband nanophotonic inverse design~\cite{lalau2013adjoint,hammond2022high, chung2000optimal,park2026multi}, organized by how the spectral content of the design gradient is stored and computed. One can always specify the objective in the frequency domain, but then the question arises of how to compute the per-frequency gradients, in a time-domain solver. At each frequency, the gradient couples forward and adjoint fields with complex weights determined by the objective, so the full broadband gradient is a spectral overlap integral of the two fields. This overlap can be evaluated in one of two ways. A \emph{frequency-domain adjoint} method evaluates it at discretely sampled frequencies, which requires the adjoint response resolved at each sample---either one adjoint simulation per frequency, or a single simulation driven by a synthesized broadband source that prescribes the required amplitude and phase at every sample. A \emph{time-domain adjoint} method instead evaluates the overlap directly in the time domain, with no spectral discretization: its adjoint simulation is driven by the objective's time-domain residual, and the frequency-resolved information is supplied by the stored forward-field history. The first route pays for spectral density in adjoint runtime; the second pays in memory. We focus on the computational cost, memory requirements, and scalability of each formulation for broadband optimization, as summarized schematically in Fig.~\ref{fig:overview}.

The four formulations in Fig.~\ref{fig:overview} expose the central runtime--memory trade-off in broadband adjoint optimization.
A per-frequency frequency-domain approach requires \(1+N_f\) simulations and stores frequency-resolved forward and adjoint fields with a dominant cost of \(2N_fN_{\mathrm{des}}\).
A synthesized-source variant restores the \(1+1\) simulation count, but its adjoint-source duration scales as \(1/\Delta f\) as the spectral spacing is refined.
Conventional time-domain adjoint optimization also retains a \(1+1\) simulation count with \(N_f\)-independent runtime, but full-history storage scales as \(2N_{\mathrm{des}}N_{\mathrm{time}}\).
Our Nyquist-sampled formulation preserves the same broadband \(1+1\) structure while reducing forward-field storage by \(N_{\mathrm{Ny}}\) and eliminating adjoint-history storage through on-the-fly accumulation, reducing the dominant memory requirement to approximately \(N_{\mathrm{des}}N_{\mathrm{time}}/N_{\mathrm{Ny}}\).

\subsection{Frequency-domain adjoint method}
\label{subsec:fda}

As illustrated in the left column of Fig.~\ref{fig:overview}, the frequency-domain adjoint method obtains the broadband forward response from a single time-domain simulation and extracts steady-state field maps at $N_f$ target frequencies using discrete Fourier transforms. In the per-frequency variant, the method then, for each frequency $f_k$, launches an adjoint simulation driven by the derivative of the objective with respect to the forward fields at that frequency. The resulting forward and adjoint fields provide the gradient of the objective with respect to all design variables at that frequency simultaneously~\cite{molesky2018inverse,lalau2013adjoint}.

Because the frequency-domain method only requires frequency-domain field profiles at discrete target frequencies, its field-storage requirement scales with the number of sampled frequencies and the size of the design region. When both
forward and adjoint field maps are retained, the memory cost scales as
\begin{equation}
  \mathcal{M}_{\text{FDA}} \propto 2\,N_f\,N_{\text{des}},
  \label{eq:mem-fda}
\end{equation}
where $N_{\text{des}}$ denotes the number of design degrees of freedom. This storage requirement remains manageable compared with storing the full-time-domain field history, even for relatively large design regions~\cite{hammond2022high}. Because of this favorable storage scaling,  discrete-Fourier-transform (DFT)-based field accumulation is the default strategy in widely used inverse-design frameworks~\cite{lalau2013adjoint,Chung:20,oskooi2010meep,hughes2021perspective,hammond2022high}, and it---rather than full time-history storage---constitutes the practical memory baseline for the comparisons that follow.

The main limitation of the frequency-domain adjoint method lies in its runtime scaling. As shown in Fig.~\ref{fig:overview}, the per-frequency variant requires one broadband forward simulation and $N_f$ frequency-specific adjoint simulations, so the total number of simulations scales as $1+N_f$. Consequently, the computational cost can become substantial for broadband objectives that require dense spectral sampling~\cite{hughes2019forward,kang2024large}. The synthesized-source variant~\cite{hammond2022high} instead uses a single adjoint simulation driven by a broadband source constructed to reproduce the desired adjoint amplitudes and phases at all sampled frequencies simultaneously. (A preliminary version of this construction, superposing adjoint excitations at a few target wavelengths without addressing the resulting source-duration cost, appears in Ref.~\citenum{mansouree2021large}.) This restores the $1+1$ simulation count, but the synthesized adjoint-source duration scales as $1/\Delta f$ and therefore increases approximately linearly with $N_f$ over a fixed bandwidth. The total adjoint simulation time need not follow the same linear scaling because it also depends on the device-dependent field-decay time and the stopping criterion. Thus, either multiple frequency-specific solves or a longer synthesized-source construction introduces an increasing runtime cost as the prescribed spectral sampling is refined.

\subsection{Time-domain adjoint method with full-history storage}
\label{subsec:tda-conv}

The conventional time-domain adjoint method, illustrated in the center column of Fig.~\ref{fig:overview}, addresses the runtime bottleneck of frequency-domain adjoint formulations by computing broadband sensitivities using only two time-domain simulations: one forward simulation and one time-reversed adjoint simulation, independent of the number of spectral samples~\cite{hughes2019forward,hassan2022topology,gedeon2023time,park2026multi}. Crucially, its adjoint excitation is not constructed to prescribe spectral amplitudes at sampled frequencies: it is simply the objective's time-domain residual, evaluated along the forward trajectory and injected time-reversed, and the broadband spectral overlap that defines the gradient is evaluated directly in the time domain. For a generic time-domain objective,
\begin{equation}
  F = \int_{\Omega_t}\!\int_{0}^{T} G\!\bigl(\mathbf{E}(\mathbf{r},t),\mathbf{H}(\mathbf{r},t)\bigr)\,dt\,d\mathbf{r},
  \label{eq:obj-td}
\end{equation}
the adjoint formulation gives the material sensitivity in the design region as
\begin{equation}
  \frac{\partial F}{\partial\rho}(\mathbf{r})
  = \sum_{n=0}^{N_{\text{time}}-1}
    \frac{\partial\varepsilon}{\partial\rho}(\mathbf{r})\,
    \mathbf{E}_{\text{fwd}}^{n}(\mathbf{r})
    \cdot
    \frac{\mathbf{E}_{\text{adj}}^{n+1}(\mathbf{r})
    - \mathbf{E}_{\text{adj}}^{n}(\mathbf{r})}{\Delta t},
  \qquad \mathbf{r}\in\Omega_D .
  \label{eq:grad-td}
\end{equation}
Here, the superscript $n$ denotes the time-step index, and the finite difference approximates the time derivative of the adjoint electric field~\cite{chung2000optimal}. The gradient expression couples the forward electric field with the time derivative of the adjoint electric field at each design voxel and time step~\cite{chung2000optimal,nomura2007structural}. Therefore, the method requires access to the forward-field history throughout the design region during the adjoint simulation. The adjoint excitation, by contrast, imposes no cost that grows with the number of frequencies. Because the objective in Eq.~\eqref{eq:obj-td} is a time-domain integral of the fields, its residual $\partial G/\partial\mathbf{E}$ exists only at the same time steps as the forward solves, and the adjoint simulation therefore comprises the same duration as the forward simulation, by construction. What sets $N_{\text{time}}$ itself is the duration of the forward simulation---the source duration plus the physical decay of the fields---independent of how densely the objective weights the spectrum. In a straightforward implementation that stores both forward and adjoint field histories for post-processing, the dominant memory cost scales as
\begin{equation}
  \mathcal{M}_{\text{TDA}} \propto 2\,N_{\text{des}}\,N_{\text{time}},
  \label{eq:mem-conv}
\end{equation}
where $N_{\text{des}}$ denotes the number of design degrees of freedom and $N_{\text{time}}$ denotes the total number of time steps. Even when the implementation accumulates the gradient on the fly and avoids storing the full adjoint history, it must still retain or reconstruct the full forward-field history, leading to a memory cost that scales as $\mathcal{O}(N_{\text{des}}N_{\text{time}})$. Thus, the conventional time-domain adjoint method offers favorable runtime scaling but imposes a severe memory burden, which becomes prohibitive for large-scale or three-dimensional inverse-design problems~\cite{kang2024large,park2025time}, as quantified in Section~\ref{subsec:memory}.

%======================================================
\section{Memory-efficient Nyquist-sampled time-domain adjoint method}
\label{sec:proposed}
%======================================================

This section introduces the proposed memory-efficient time-domain adjoint framework. The method preserves the two-simulation structure of conventional time-domain adjoint optimization while reducing the dominant forward-field storage requirement through Nyquist-sampled temporal downsampling. We first describe how the bandwidth-limited nature of FDTD field responses enables sparse storage of forward fields without sacrificing the information required for gradient evaluation. We then explain how the adjoint simulation accumulates the gradient on the fly using the stored field representation, thereby avoiding the need to retain the full adjoint-field history. Together, these two components substantially reduce memory consumption (from the FDTD step count $N_{\text{time}}$ to the information-theoretic floor set by the field bandwidth) while maintaining the scalability ($N_f$-independent runtimes) and broadband capability of time-domain adjoint optimization. 

\subsection{Nyquist-sampled forward-field storage}
\label{subsec:nyquist}

The proposed method, illustrated in the right column of Fig.~\ref{fig:overview}, retains the $1+1$ simulation count of the conventional time-domain adjoint formulation while substantially reducing its memory consumption. The key idea is to store the forward fields at the minimum temporal sampling rate required to represent the bandwidth-limited electromagnetic response, rather than at every FDTD time step. Let $u(t)$ denote a representative time trace of a field component at a point in the design region. If the spectral content of $u(t)$ is confined to $|f|\le f_{\max}$, the Nyquist--Shannon sampling theorem guarantees that $u(t)$ can be reconstructed from uniformly spaced samples with a sampling interval $\Delta t_s \le (2f_{\max})^{-1}$ in the ideal band-limited limit~\cite{shannon1949communication}. Since the FDTD time step $\Delta t$ is typically much smaller than this Nyquist interval because of the CFL stability condition and spatial-resolution requirements, the full forward-field history contains substantial temporal redundancy. We exploit this redundancy by storing the forward fields only every $N_{\mathrm{Ny}}$ FDTD time steps,
\begin{equation}
  \Delta t_s = N_{\mathrm{Ny}}\Delta t,
  \qquad
  u_s[m] = u(mN_{\mathrm{Ny}}\Delta t),
  \label{eq:down-sample}
\end{equation}
where the integer downsampling factor is chosen as
\begin{equation}
  N_{\mathrm{Ny}} =
  \max\left(1,\left\lfloor\frac{1}{2f_{\max}\Delta t}\right\rfloor\right).
  \label{eq:Nny}
\end{equation}

This choice ensures that the stored sampling interval does not exceed the Nyquist limit. In practice, $f_{\max}$ should cover the source bandwidth and the spectral range over which the objective and adjoint response contribute to the gradient, preferably with a small safety margin to suppress aliasing from weak high-frequency components. During the adjoint simulation, the method evaluates the overlap integral directly on the downsampled temporal grid: contributions are accumulated only at the stored time steps, with the coarse-grid quadrature weight $\Delta t_s = N_{\mathrm{Ny}}\Delta t$, and no interpolation or reconstruction to the full FDTD rate is required. Thus, instead of storing the complete forward and adjoint field histories, the proposed method stores only the downsampled forward-field history and retains only the instantaneous adjoint fields required for gradient accumulation. The gradient can be written as
\begin{equation}
  \frac{\partial F}{\partial\rho}(\mathbf{r})
  =
  {N_{\mathrm{Ny}}\!\sum_{m=0}^{\lceil N_{\text{time}}/N_{\mathrm{Ny}}\rceil-1}}
  \frac{\partial\varepsilon}{\partial\rho}(\mathbf{r})\,
  {\mathbf{E}_{\text{fwd}}^{\,mN_{\mathrm{Ny}}}(\mathbf{r})}
  \cdot
  \frac{\mathbf{E}_{\text{adj}}^{{mN_{\mathrm{Ny}}}+1}(\mathbf{r})
  - \mathbf{E}_{\text{adj}}^{{mN_{\mathrm{Ny}}}}(\mathbf{r})}{\Delta t},
  \qquad \mathbf{r}\in\Omega_D,
  \label{eq:grad-desa}
\end{equation}
where the sum runs over the stored samples and the prefactor $N_{\mathrm{Ny}}$ is the quadrature weight of the coarse temporal grid relative to Eq.~\eqref{eq:grad-td}. Under the assumed band-limited condition, this representation preserves the forward-field information within the target bandwidth, thereby maintaining the accuracy of the adjoint gradient. The dominant memory cost becomes
\begin{equation}
  \mathcal{M}_{\text{prop}}
  \propto
  N_{\text{des}}
  \left\lceil
  \frac{N_{\text{time}}}{N_{\mathrm{Ny}}}
  \right\rceil
  + \mathcal{O}(N_{\text{des}}),
  \label{eq:mem-desa}
\end{equation}
where the first term accounts for the stored downsampled forward fields and the second term accounts for the instantaneous adjoint fields used during on-the-fly accumulation. Compared with a conventional implementation that stores the full forward-field history, the proposed method reduces the dominant storage requirement by approximately a factor of $N_{\mathrm{Ny}}$ while preserving the two-simulation runtime advantage of the time-domain adjoint method. The stored representation then comprises $N_{\text{time}}/N_{\mathrm{Ny}} \approx 2f_{\max}T$ samples per field component---precisely the number of independent degrees of freedom carried by a signal of duration $T$ and bandwidth $f_{\max}$. Nyquist-sampled storage is thus informationally equivalent to retaining frequency-domain field maps at the full spectral resolution $1/T$ of the simulation across the entire band: it stores the complete spectral content of the forward field at the minimum cost the sampling theorem allows.

\subsection{On-the-fly adjoint-gradient accumulation}
\label{subsec:onthefly}

We accumulate the gradient on the fly during the adjoint simulation to avoid storing the full adjoint-field history. At each adjoint time step that coincides with a stored forward sample, $n = mN_{\mathrm{Ny}}$, the method retrieves the stored forward field and immediately updates the gradient accumulator as
\begin{equation}
  \frac{\partial F}{\partial\rho}(\mathbf{r})
  \leftarrow
  \frac{\partial F}{\partial\rho}(\mathbf{r})
  +
  \frac{\partial\varepsilon}{\partial\rho}(\mathbf{r})\,
  {\mathbf{E}_{\mathrm{fwd}}^{\,mN_{\mathrm{Ny}}}(\mathbf{r})}
  \cdot
  \frac{\mathbf{E}_{\mathrm{adj}}^{{mN_{\mathrm{Ny}}}+1}(\mathbf{r})
  - \mathbf{E}_{\mathrm{adj}}^{{mN_{\mathrm{Ny}}}}(\mathbf{r})}{\Delta t}\,
  {N_{\mathrm{Ny}}}\Delta t,
  \qquad \mathbf{r}\in\Omega_D .
  \label{eq:onthefly}
\end{equation}

{At adjoint time steps between stored samples, no accumulation is performed. Because the update uses only the instantaneous adjoint fields and the stored forward field, the method eliminates the need to store the adjoint-field history. Note that the band-limit argument requires care here, because the accumulation samples the \emph{product} of forward and adjoint fields, whose spectral content extends to $2f_{\max}$---twice the band limit of either factor. The time integral is nevertheless evaluated exactly: by Poisson summation, a discrete sum at sampling rate $\Delta t_s^{-1} = 2f_{\max}$ reproduces the integral of any signal whose spectrum vanishes at nonzero integer multiples of the sampling rate, and the product's spectrum, confined below $2f_{\max}$, satisfies this condition.} The memory footprint is therefore dominated by the downsampled forward-field buffer and a single gradient accumulator, with the scaling given by Eq.~\eqref{eq:mem-desa}.

Compared with a conventional implementation that stores both forward and adjoint field histories, this scaling corresponds to an approximate memory reduction factor of $2N_{\mathrm{Ny}}$. Compared with an implementation that already accumulates the adjoint contribution on the fly but still stores the complete forward-field history, the reduction factor is approximately $N_{\mathrm{Ny}}$. A further consequence of coarse-grid accumulation is that the gradient-update arithmetic and its memory traffic also drop by the factor $N_{\mathrm{Ny}}$, so the adjoint pass can be marginally faster than a full-storage implementation, consistent with the per-iteration runtimes reported in Section~\ref{sec:results}.

%================================
\subsection{Algorithm summary and implementation}
\label{subsec:algo}
%================================

Algorithms~\ref{alg:conventional_adjoint} and~\ref{alg:nyquist_adjoint} summarize the conventional and proposed time-domain adjoint workflows, respectively, and clarify their differences in field storage and gradient evaluation. In a straightforward conventional implementation, the method stores the forward and adjoint fields throughout the design region at every time step and evaluates the gradient in a separate post-processing step. This strategy requires a memory consumption that scales as $2N_{\text{time}}N_{\text{des}} + N_{\text{FDTD}}$, where $N_{\text{FDTD}}$ denotes the memory required for the instantaneous FDTD field arrays over the computational domain. In contrast, the proposed method stores only the Nyquist-sampled forward-field history and accumulates the gradient during the adjoint simulation. This implementation reduces the dominant storage requirement to approximately $(N_{\text{time}}/N_{\mathrm{Ny}})N_{\text{des}} + N_{\text{FDTD}} + \mathcal{O}(N_{\text{des}})$, where the last term accounts for the gradient accumulator. Compared with the conventional full-history implementation, this corresponds to an approximate memory reduction factor of $2N_{\mathrm{Ny}}$ for the design-region field storage. Importantly, the proposed strategy modifies only the data-storage and gradient-accumulation procedure; it leaves the underlying FDTD update equations unchanged~\cite{yee1966numerical,taflove2005computational}.

\begin{algorithmblock}[alg:conventional_adjoint]
{Conventional time-domain adjoint method with full field-history storage.}
The method stores the forward and adjoint electric fields at every time step
throughout the design region and evaluates the gradient in a post-processing
step. This full-storage implementation has a dominant memory cost of
$2N_{\mathrm{time}}N_{\mathrm{des}} + N_{\mathrm{FDTD}}$.

\begin{algorithmic}[1]
\Require Design parameters $\boldsymbol{\rho}$, time step $\Delta t$,
total number of time steps $N_{\mathrm{time}}$
\Ensure Gradient $\partial {F}/\partial\boldsymbol{\rho}$
\Statex
\State \textbf{Forward simulation with full storage}
\State Initialize the forward FDTD fields.
\For{$n = 0$ \textbf{to} $N_{\mathrm{time}}-1$}
  \State Advance Maxwell's equations by one time step.
  \State Store $\mathbf{E}_{\mathrm{fwd}}^{n}(\mathbf{r})$
  for all $\mathbf{r}\in\Omega_D$.
  \State Record the monitor fields required to evaluate the objective $F$.
\EndFor
\State Evaluate the objective ${F}$ and construct the corresponding adjoint source.
\Statex
\State \textbf{Adjoint simulation with full storage}
\State Initialize the adjoint FDTD fields.
\For{$n = N_{\mathrm{time}}-1$ \textbf{downto} $0$}
  \State Advance the adjoint Maxwell equations by one time step in reverse time.
  \State Store $\mathbf{E}_{\mathrm{adj}}^{n}(\mathbf{r})$
  for all $\mathbf{r}\in\Omega_D$.
\EndFor
\Statex
\State \textbf{Gradient computation by post-processing}
\State Initialize $\partial {F}/\partial\boldsymbol{\rho} \gets \mathbf{0}$.
\For{$n = 0$ \textbf{to} $N_{\mathrm{time}}-1$}
  \State $\displaystyle
    \frac{\partial {F}}{\partial\boldsymbol{\rho}}
    \gets
    \frac{\partial {F}}{\partial\boldsymbol{\rho}}
    +
    \frac{\partial\varepsilon}{\partial\boldsymbol{\rho}}\,
    \mathbf{E}_{\mathrm{fwd}}^{n}
    \cdot
    \frac{\mathbf{E}_{\mathrm{adj}}^{n+1}
    - \mathbf{E}_{\mathrm{adj}}^{n}}{\Delta t}$
\EndFor
\State \Return $\partial {F}/\partial\boldsymbol{\rho}$
\Statex
\State \textbf{Memory cost:}
$2N_{\mathrm{time}}N_{\mathrm{des}} + N_{\mathrm{FDTD}}$
\end{algorithmic}
\end{algorithmblock}

\begin{algorithmblock}[alg:nyquist_adjoint]
{Proposed Nyquist-sampled time-domain adjoint method.}
The forward simulation stores the electric fields in the design region only
at Nyquist-sampled time intervals determined by $N_{\mathrm{Ny}}$.
During the adjoint simulation, the method accumulates the gradient only at the stored time steps.

\begin{algorithmic}[1]
\Require Design parameters $\boldsymbol{\rho}$, maximum relevant frequency
$f_{\max}$, time step $\Delta t$, total number of time steps $N_{\mathrm{time}}$
\Ensure Gradient $\partial {F}/\partial\boldsymbol{\rho}$
\Statex
\State \textbf{Compute the Nyquist downsampling factor}
\State $N_{\mathrm{Ny}}
\gets
\max\left(1,\left\lfloor 1/(2f_{\max}\Delta t)\right\rfloor\right)$
\Comment{$\Delta t_s=N_{\mathrm{Ny}}\Delta t \le 1/(2f_{\max})$}
\Statex
\State \textbf{Forward simulation with Nyquist-sampled storage}
\State Initialize the forward FDTD fields.
\For{$n = 0$ \textbf{to} $N_{\mathrm{time}}-1$}
  \State Advance Maxwell's equations by one time step.
  \State Record the monitor fields required to evaluate the objective $F$.
  \If{$n \bmod N_{\mathrm{Ny}} = 0$}
    \State Store $\mathbf{E}_{\mathrm{fwd}}^{n}(\mathbf{r})$
    for all $\mathbf{r}\in\Omega_D$.
  \EndIf
\EndFor
\State Evaluate the objective ${F}$ and construct the corresponding adjoint source.
\Statex
\State \textbf{Adjoint simulation with on-the-fly gradient accumulation}
\State Initialize $\partial {F}/\partial\boldsymbol{\rho} \gets \mathbf{0}$.
\State Initialize the adjoint FDTD fields.
\For{$n = N_{\mathrm{time}}-1$ \textbf{downto} $0$}
  \State Advance the adjoint Maxwell equations by one reverse-time step.
  {
  \If{$n \bmod N_{\mathrm{Ny}} = 0$}
  \State $\displaystyle
    \frac{\partial {F}}{\partial\boldsymbol{\rho}}
    \gets
    \frac{\partial {F}}{\partial\boldsymbol{\rho}}
    +
    \frac{\partial\varepsilon}{\partial\boldsymbol{\rho}}\,
    \mathbf{E}_{\mathrm{fwd}}^{\,n}
    \cdot
    \frac{\mathbf{E}_{\mathrm{adj}}^{n+1}
    - \mathbf{E}_{\mathrm{adj}}^{n}}{\Delta t}\,N_{\mathrm{Ny}}$
  \EndIf
  }
\EndFor
\State \Return $\partial {F}/\partial\boldsymbol{\rho}$
\Statex
\State \textbf{Memory cost:}
$\left\lceil N_{\mathrm{time}}/N_{\mathrm{Ny}}\right\rceil
N_{\mathrm{des}} + N_{\mathrm{FDTD}} + \mathcal{O}(N_{\mathrm{des}})$
\Comment{Approximately $2N_{\mathrm{Ny}}$ reduction relative to
Algorithm~\ref{alg:conventional_adjoint}.}
\end{algorithmic}
\end{algorithmblock}

% ============================================================
\section{Results and Discussion}
\label{sec:results}
% ============================================================

This section evaluates the proposed Nyquist-sampled time-domain adjoint method from three complementary perspectives. First, we quantify measured memory and runtime performance using representative two-dimensional benchmarks; extended three-dimensional memory-scaling projections are provided in the Supporting Information. Second, we examine the adjoint-gradient accuracy as a function of the downsampling factor $N_{\mathrm{Ny}}$, including the transition beyond the Nyquist limit. Third, we demonstrate the method on broadband inverse-designed devices in two and three dimensions, including a fully simulated three-dimensional metalens. We performed all adjoint optimizations using an in-house FDTD solver and cross-validated selected results against the open-source Meep FDTD package~\cite{oskooi2010meep}.

%==============================
\subsection{Memory scaling of Nyquist-sampled adjoint optimization}
\label{subsec:memory}
%==============================

Table~\ref{tab:2d_memory} summarizes the measured memory and runtime
requirements of the conventional full-storage and proposed Nyquist-sampled
time-domain adjoint methods for representative two-dimensional benchmarks.
For both methods, the dominant field-storage memory scales approximately
linearly with the number of design cells, but the proposed method
substantially reduces the proportionality constant through Nyquist-sampled
forward-field storage and on-the-fly gradient accumulation.

\begin{table*}[t]
\centering
\caption{
Memory and runtime comparison between conventional full-storage and proposed
Nyquist-sampled time-domain adjoint(TDA) methods for representative 2D
nanophotonic benchmarks. The listed temporal parameters specify the
Nyquist-sampling conditions used in each case.
}
\label{tab:2d_memory}

\scriptsize
\setlength{\tabcolsep}{2.4pt}
\renewcommand{\arraystretch}{0.92}

\resizebox{\textwidth}{!}{%
\begin{tabular}{llcccccccccc}
\toprule
Application
& Configuration
& \shortstack{Comput.\\domain\\(\# cells)}
& \shortstack{Design\\region\\(\# cells)}
& $N_{\mathrm{time}}$
& \shortstack{$\Delta t$\\(fs)}
& \shortstack{$\lambda_{\min}$\\(nm)}
& $N_{\mathrm{Ny}}^{\max}$
& $N_{\mathrm{Ny}}$
& \shortstack{Conventional TDA\\Time (s) / Mem. (GB)}
& \shortstack{\textbf{Proposed Nyquist-TDA}\\Time (s) / Mem. (GB)}
& \shortstack{Memory\\reduction} \\
\midrule

\multirow{2}{*}{Metalens (NA\,0.6)}
& Standard
& 208,000 & 10,000
& 6,000
& 0.02335
& 490
& 34
& \textbf{34}
& 7.91 / 1.62
& 7.21 / \textbf{0.02}
& $\approx 81\times$ \\
& Memory-limited
& 208,000 & 10,000
& 240,000
& 0.02335
& 490
& 34
& \textbf{34}
& \textbf{926.31$^\ddagger$} / \textbf{131.5$^\dagger$}
& -- / \textbf{5.58}
& $\approx 23\times$ \\
\midrule

\multirow{2}{*}{Wavelength demux}
& Standard
& 176,400 & 40,000
& 8,000
& 0.02335
& 1000
& 71
& \textbf{71}
& 8.02 / 6.43
& 7.22 / \textbf{0.06}
& $\approx 107\times$ \\
& Memory-limited
& 176,400 & 40,000
& 208,000
& 0.02335
& 1000
& 71
& \textbf{71}
& \textbf{917.48$^\ddagger$} / \textbf{167.5$^\dagger$}
& -- / \textbf{1.34}
& $\approx 125\times$ \\
\midrule

\multirow{2}{*}{Waveguide bend}
& Standard
& 384,400 & 22,500
& 8,000
& 0.02335
& 1000
& 71
& \textbf{71}
& 14.21 / 2.99
& 13.54 / \textbf{0.03}
& $\approx 100\times$ \\
& Memory-limited
& 384,400 & 22,500
& 160,000
& 0.02335
& 1000
& 71
& \textbf{71}
& \textbf{1231.79$^\ddagger$} / \textbf{162.2$^\dagger$}
& -- / \textbf{1.41}
& $\approx 115\times$ \\
\midrule

\multirow{2}{*}{Color router}
& Standard
& 74,000 & 60,000
& 12,000
& 0.02335
& 400
& 28
& \textbf{24}
& 5.71 / 11.90
& 4.10 / \textbf{0.16}
& $\approx 74\times$ \\
& Memory-limited
& 74,000 & 60,000
& 360,000
& 0.02335
& 400
& 28
& \textbf{24}
& \textbf{823.21$^\ddagger$} / \textbf{184.8$^\dagger$}
& -- / \textbf{4.03}
& $\approx 46\times$ \\
\bottomrule
\end{tabular}%
}

\vspace{2pt}
\parbox{\textwidth}{%
\footnotesize
\textit{Note:}
All benchmarks use $\Delta x=\Delta y=10~\mathrm{nm}$,
$\mathrm{CFL}=0.99$, and $\Delta t=0.02335~\mathrm{fs}$.
$N_{\mathrm{Ny}}^{\max}$ is evaluated from the Nyquist condition for the
listed $\lambda_{\min}$.
For the color router, $N_{\mathrm{Ny}}=24$ was chosen below
$N_{\mathrm{Ny}}^{\max}=28$ to retain a safety margin from the Nyquist limit.
$\dagger$ denotes configurations exceeding the available 128~GB memory;
$\ddagger$ denotes runtimes measured under memory-saturated conditions.
A dash (--) denotes an unreported runtime.
}

\end{table*}

Under the standard configuration, the proposed method reduces the
field-storage memory by approximately one to two orders of magnitude while
maintaining a per-iteration runtime comparable to, or slightly lower than,
that of the conventional full-storage implementation.
The largest standard-case reduction occurs for the wavelength demultiplexer,
where the memory requirement decreases from 6.43~GB to 0.06~GB,
corresponding to approximately \(107\times\).

The distinction becomes more pronounced in the memory-limited cases.
Conventional full-history storage exceeds the available 128~GB system memory,
leading to paging and out-of-core data movement and a substantial runtime
increase.
By contrast, the proposed method remains within the available memory without
recomputing the intervening forward fields.
Checkpointing can also reduce memory in this regime, but does so by trading
storage for additional forward-field recomputation~\cite{griewank2000algorithm,wang2009minimal}.
Thus, Nyquist-sampled storage directly removes temporally redundant field
samples while preserving the two-simulation structure of time-domain
adjoint optimization.

Extended memory-scaling projections for representative three-dimensional
devices and a large-area metalens are provided in Section~S1
(Fig.~S1) of the Supporting Information.

%==============================
\subsection{Gradient accuracy under Nyquist-sampled field storage}
\label{subsec:gradient}
%==============================

\begin{figure}[!htbp]
  \centering
  \includegraphics[width=\textwidth]{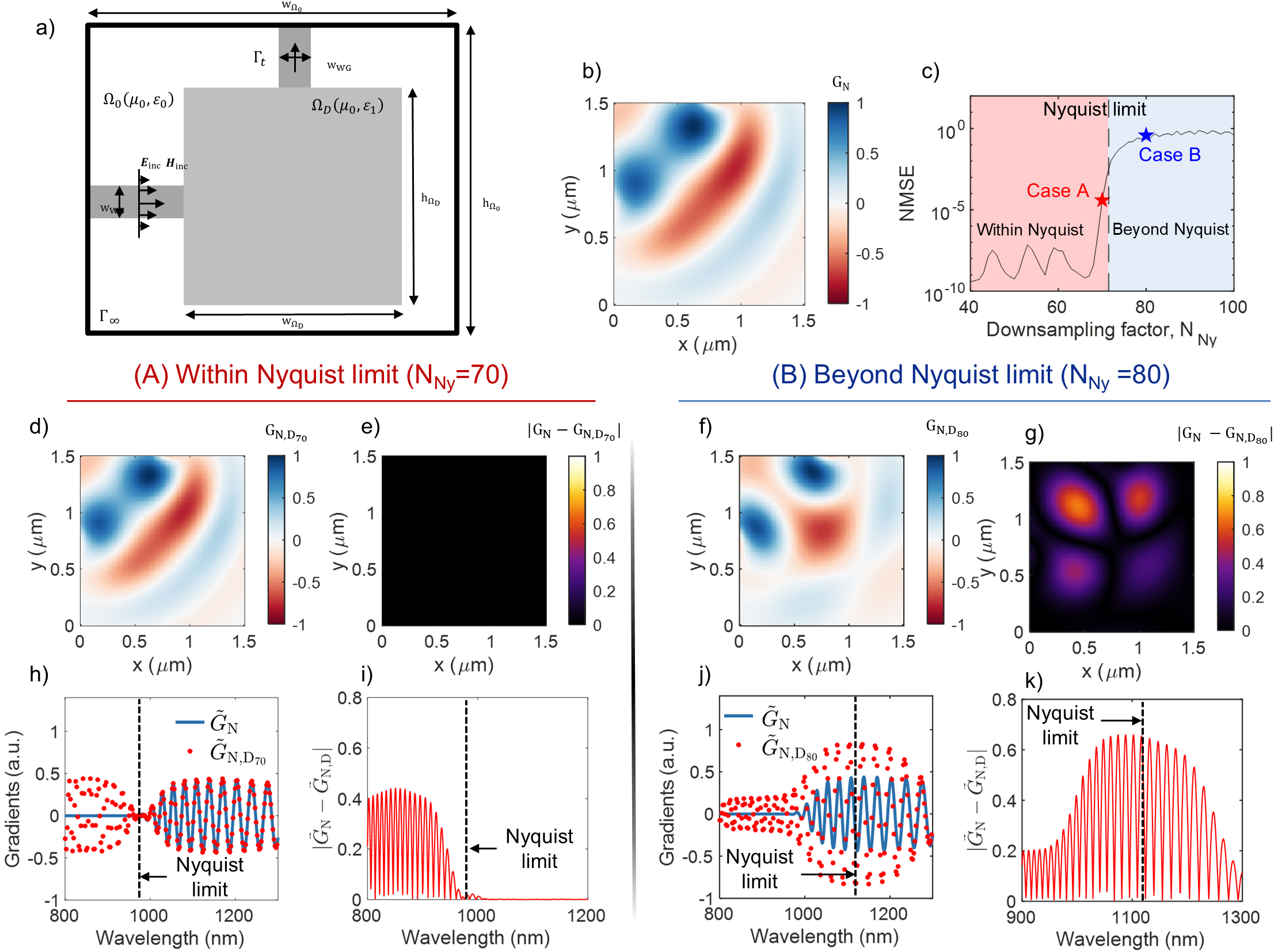}
\caption{
Gradient-accuracy verification of the proposed Nyquist-sampled
time-domain adjoint method.
(a) Verification geometry: a
$6\,\mu\mathrm{m}\times6\,\mu\mathrm{m}$ domain
$\Omega_0(\mu_0,\varepsilon_0)$ with a
$1.5\,\mu\mathrm{m}\times1.5\,\mu\mathrm{m}$ design region
$\Omega_D(\mu_0,\varepsilon_1)$, where
$\varepsilon_1=12.11$ and $\Delta x=\Delta y=10\,\mathrm{nm}$.
A broadband guided source is injected through a
$0.3\,\mu\mathrm{m}$-wide input waveguide, and the objective is evaluated at
the top output waveguide $\Gamma_t$.
(b) Reference normalized gradient map $G_N$ from the conventional
full-storage time-domain adjoint method.
(c) NMSE between $G_N$ and the downsampled gradient versus the downsampling
factor $N_{\mathrm{Ny}}$; the dashed line denotes the Nyquist limit for
$\lambda_{\min}=1000\,\mathrm{nm}$.
Case~A ($N_{\mathrm{Ny}}=70$) lies within the limit, whereas Case~B
($N_{\mathrm{Ny}}=80$) exceeds it.
(d,e) Downsampled gradient $G_{N,D_{70}}$ and error
$|G_N-G_{N,D_{70}}|$ for Case~A.
(f,g) Corresponding results for Case~B, showing pronounced spatial error.
(h,i) Spectral comparison and residual for Case~A.
(j,k) Spectral comparison and residual for Case~B, showing aliasing-induced
deviation beyond the Nyquist limit.
These results confirm that Nyquist-compliant downsampling preserves
adjoint-gradient fidelity, whereas undersampling degrades both spatial and
spectral accuracy.
}
  \label{fig:gradient}
\end{figure}

We verify the accuracy of the adjoint gradient obtained with Nyquist-sampled
forward-field storage using the two-dimensional waveguide benchmark shown in
Fig.~\ref{fig:gradient}(a).
The computational domain $\Omega_0$ spans
$6\,\mu\mathrm{m}\times6\,\mu\mathrm{m}$ and contains a
$1.5\,\mu\mathrm{m}\times1.5\,\mu\mathrm{m}$ design region $\Omega_D$ with
relative permittivity $\varepsilon_1=12.11$, discretized on a uniform
Cartesian grid with $\Delta x=\Delta y=10\,\mathrm{nm}$.
A broadband guided source excites the structure from the left through a
$0.3\,\mu\mathrm{m}$-wide input waveguide, and the objective evaluates the
field intensity collected at the top output waveguide~\cite{chung2000optimal}.
The transverse-magnetic field components $H_x$, $H_y$, and $E_z$ are placed
on the standard two-dimensional Yee lattice~\cite{yee1966numerical}.

Figure~\ref{fig:gradient} summarizes the gradient-validation results.
The normalized reference gradient map $G_N$ in Fig.~\ref{fig:gradient}(b)
is obtained using full time-step storage.
Figure~\ref{fig:gradient}(c) shows the normalized mean-square error (NMSE) between this reference gradient
and the gradient computed using downsampled forward-field storage as a
function of the downsampling factor $N_{\mathrm{Ny}}$.
The vertical dashed line marks the Nyquist limit determined by the highest
frequency in the target band, corresponding here to
$\lambda_{\min}=1000\,\mathrm{nm}$.
Within the Nyquist-compliant region, the NMSE remains near the numerical
noise floor, indicating that the downsampled field history retains the
information required for accurate gradient recovery.
Once $N_{\mathrm{Ny}}$ exceeds the Nyquist limit, the NMSE increases
abruptly by several orders of magnitude and reaches order-unity values,
signaling severe degradation of the reconstructed gradient.

We examine two representative cases in detail.
Case~A uses $N_{\mathrm{Ny}}=70$, which lies within the Nyquist limit.
The downsampled gradient map $G_{N,D_{70}}$ in
Fig.~\ref{fig:gradient}(d) is visually indistinguishable from the reference
map in Fig.~\ref{fig:gradient}(b), and the corresponding absolute error map
in Fig.~\ref{fig:gradient}(e) remains nearly zero throughout the design
region.
The same conclusion is obtained in the wavelength domain:
the reference gradient spectrum $\tilde{G}_N$ and the downsampled spectrum
$\tilde{G}_{N,D_{70}}$ overlap closely in Fig.~\ref{fig:gradient}(h), while
the residual in Fig.~\ref{fig:gradient}(i) remains negligible within the
properly sampled spectral range.
Thus, when the downsampling factor satisfies the Nyquist criterion, the
proposed method reproduces the conventional adjoint gradient with no
appreciable loss of accuracy while reducing the number of stored temporal
samples by approximately $70\times$.
Case~A sits only just below the Nyquist boundary itself, yet its NMSE already lies at the numerical noise floor: within these benchmarks, the band-limited assumption holds sharply enough that only a small safety margin below the boundary is needed.

By contrast, Case~B uses $N_{\mathrm{Ny}}=80$, which exceeds the Nyquist limit.
The downsampled gradient map $G_{N,D_{80}}$ in Fig.~\ref{fig:gradient}(f) shows clear deviations from the reference, with spatially structured errors in Fig.~\ref{fig:gradient}(g).
The corresponding spectral comparison in Fig.~\ref{fig:gradient}(j) and 
residual in Fig.~\ref{fig:gradient}(k) likewise show substantial
aliasing-induced errors.
These results establish the Nyquist limit as the practical boundary for
accurate gradient recovery: once it is exceeded, aliasing in the coarse-grid
accumulation prevents the discrete gradient sum from reproducing the
full-time-step result (Section~\ref{subsec:onthefly}).

%==============================
\subsection{Nanophotonic device demonstrations}
\label{subsec:devices}
%==============================

We further validate the proposed method using five broadband photonic
inverse-design benchmarks of increasing complexity: a 2D waveguide bend, a
2D metalens, a 2D wavelength demultiplexer, a 2D color router, and a 3D
metalens.
All devices were optimized using the proposed Nyquist-sampled time-domain
adjoint method.
The gradient-verification results in Section~\ref{subsec:gradient} show that
Nyquist-compliant forward-field downsampling preserves the adjoint
sensitivity with an NMSE below 0.1\%.
Therefore, the proposed method provides the same optimization direction as
the conventional time-domain adjoint formulation within numerical accuracy,
while substantially reducing the memory required for field
storage~\cite{chung2000optimal,hassan2022topology}.
These device demonstrations assess whether this memory reduction translates
into practical broadband inverse-design performance across both two- and
three-dimensional photonic structures.

The downsampling factors were selected relative to the continuous Nyquist-limiting value
\[
N_{\mathrm{Ny}}^{\mathrm{lim}}
=
\frac{1}{2f_{\max}\Delta t},
\]
with the largest admissible integer factor defined as
\(N_{\mathrm{Ny}}^{\max}=\lfloor N_{\mathrm{Ny}}^{\mathrm{lim}}\rfloor\).
For the 2D devices, \(\Delta x=\Delta y=10~\mathrm{nm}\),
\(\mathrm{CFL}=0.99\), and \(\Delta t=0.02335~\mathrm{fs}\).
For the 2D metalens with \(\lambda_{\min}=490~\mathrm{nm}\),
\(N_{\mathrm{Ny}}^{\mathrm{lim}}=34.99\) and
\(N_{\mathrm{Ny}}^{\max}=34\); we use \(N_{\mathrm{Ny}}=34\),
corresponding to a margin of approximately \(0.99\) FDTD step below the
continuous Nyquist limit.
For the near-infrared devices with \(\lambda_{\min}=1000~\mathrm{nm}\),
\(N_{\mathrm{Ny}}^{\mathrm{lim}}=71.43\) and
\(N_{\mathrm{Ny}}^{\max}=71\); we use \(N_{\mathrm{Ny}}=71\),
leaving a margin of approximately \(0.43\) FDTD steps below the continuous
Nyquist limit. For the color router with \(\lambda_{\min}=400~\mathrm{nm}\),
\(N_{\mathrm{Ny}}^{\mathrm{lim}}=28.57\) and
\(N_{\mathrm{Ny}}^{\max}=28\); we use \(N_{\mathrm{Ny}}=24\),
leaving a margin of four integer steps below the largest admissible
downsampling factor.
For the 3D metalens, \(\Delta x=\Delta y=\Delta z=10~\mathrm{nm}\) gives
\(\Delta t=0.01905~\mathrm{fs}\).
For \(\lambda_{\min}=440~\mathrm{nm}\),
\(N_{\mathrm{Ny}}^{\mathrm{lim}}=38.52\) and
\(N_{\mathrm{Ny}}^{\max}=38\).
The 3D optimization uses \(N_{\mathrm{Ny}}=37\), leaving one integer step below the largest admissible factor. For the time-domain field-plane captures, we use \(N_{\mathrm{Ny}}=34\), leaving a four-step margin.

% == 2D Waveguide bend ==
\subsubsection{2D waveguide bend}
\label{subsubsec:wgbend}

\begin{figure}[!htbp]
  \centering
  \includegraphics[width=\textwidth]{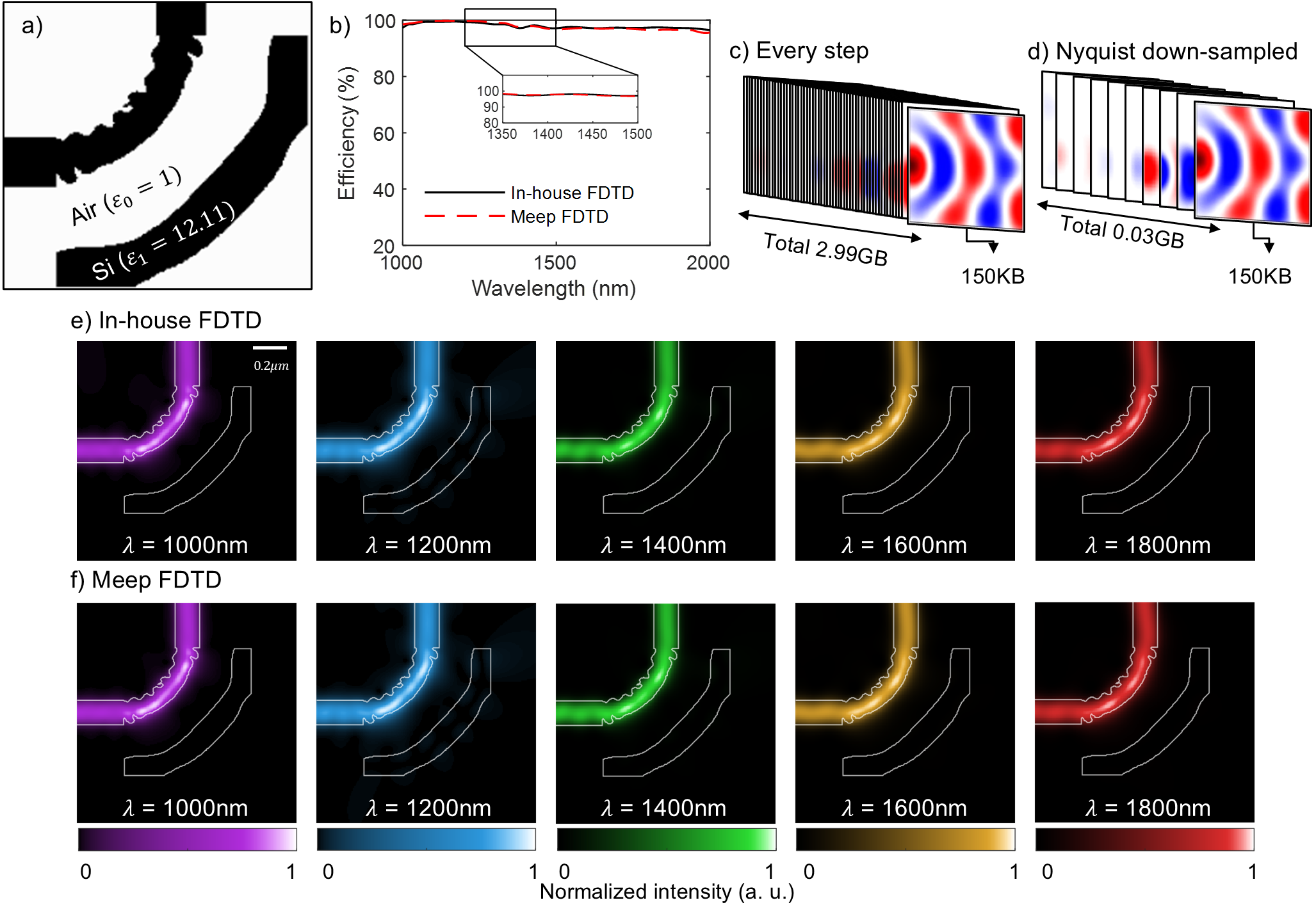}
  \caption{
  Broadband two-dimensional waveguide-bend demonstration.
  (a) Optimized freeform Si/air waveguide-bend geometry with
  $\varepsilon_0=1$ for air and $\varepsilon_1=12.11$ for silicon.
  (b) Transmission efficiency spectrum over the 1000--2000~nm wavelength
  range, cross-validated between the in-house FDTD solver and Meep
  FDTD~\cite{oskooi2010meep}.
  The inset magnifies the 1350--1500~nm region, showing close agreement
  between the two solvers.
  (c,d) Memory comparison between conventional every-step field storage and
  the proposed Nyquist-sampled field storage.
  The conventional implementation stores 150~KB field snapshots at every
  FDTD time step, resulting in 2.99~GB of forward-field storage, whereas
  the proposed method stores the same 150~KB snapshots only at
  Nyquist-sampled temporal intervals, reducing the storage to 0.03~GB.
  (e,f) Normalized electric-field intensity $|\mathbf{E}|^2$ distributions
  in and around the optimized bend at representative wavelengths
  $\lambda = 1000$, 1200, 1400, 1600, and 1800~nm.
  Panels (e) and (f) show results obtained using the in-house FDTD solver
  and Meep FDTD, respectively.
  White outlines indicate the device boundaries, and the color bars denote
  the normalized intensity scale for each wavelength.
  }
  \label{fig:bend}
\end{figure}

As the first device benchmark, we optimized a two-dimensional waveguide bend
using the proposed Nyquist-sampled time-domain adjoint
method~\cite{piggott2015inverse,piggott2020inverse}.
The device uses a $1.5\,\mu\mathrm{m}\times1.5\,\mu\mathrm{m}$ Si/air
design region with silicon relative permittivity $\varepsilon_1=12.11$,
following the benchmark configuration described in
Section~\ref{subsec:gradient}.
The optimized geometry, broadband transmission response, memory reduction,
and cross-solver field validation are summarized in Fig.~\ref{fig:bend}.

The optimized freeform silicon topology, shown in Fig.~\ref{fig:bend}(a),
routes broadband light from the left input waveguide to the top output
waveguide within a compact footprint.
The geometry obtained with the proposed method matches the design produced
by the conventional full-storage time-domain adjoint method under the same
optimization conditions, confirming that Nyquist-compliant forward-field
sampling does not degrade the final optimized structure.
The broadband transmission spectrum in Fig.~\ref{fig:bend}(b) shows that
the optimized bend maintains a transmission efficiency exceeding 95\% over
the target wavelength range.
The in-house FDTD solver and the open-source Meep FDTD package show
excellent agreement across the full spectrum, as further emphasized by the
magnified inset.

The memory reduction achieved by the proposed method is quantified in
Figs.~\ref{fig:bend}(c) and~\ref{fig:bend}(d).
In the conventional full-storage implementation, the forward simulation
stores a 150~KB field snapshot at each FDTD time step, resulting in a total
forward-field storage requirement of 2.99~GB.
In contrast, the proposed method stores the same 150~KB snapshot only at
Nyquist-sampled temporal intervals, reducing the total storage to 0.03~GB.
This corresponds to a $100\times$ reduction in field storage memory while
preserving the broadband optimization result and the two-simulation
structure of the time-domain adjoint method.

The broadband field profiles in Figs.~\ref{fig:bend}(e) and
\ref{fig:bend}(f) further verify the operation of the optimized bend using
normalized electric-field intensity maps at five representative wavelengths:
$\lambda = 1000$, 1200, 1400, 1600, and 1800~nm.
The top and bottom rows show results from the in-house FDTD solver and
Meep, respectively.
Both solvers produce nearly identical field distributions over the
representative wavelength range.
The fields remain well confined along the bend trajectory with negligible
radiation leakage, demonstrating that the proposed memory-efficient adjoint
framework can produce high-performance broadband devices while substantially
reducing the field-storage burden.
% == 2D Metalens ==
\subsubsection{2D metalens}
\label{subsubsec:metalens2d}

\begin{figure}
  \centering
  \includegraphics[width=\textwidth]{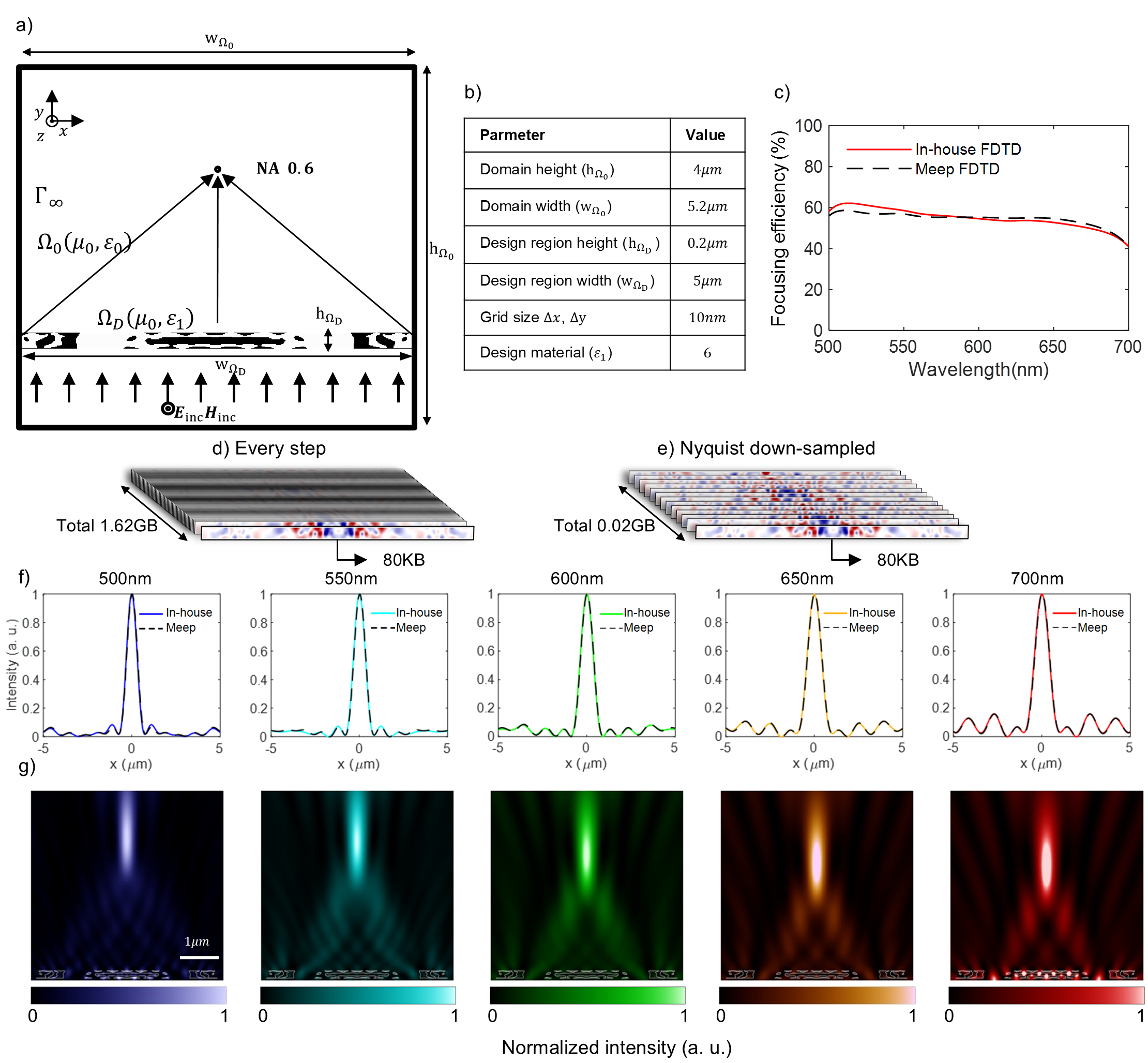}
\caption{
Broadband two-dimensional metalens demonstration. 
(a) Schematic of the optimized metalens with $\mathrm{NA}=0.6$ under normally incident broadband illumination. 
(b) Key simulation and design parameters. 
(c) Focusing efficiency spectrum over 500--700 nm, cross-validated between the in-house FDTD solver and Meep FDTD. 
(d,e) Memory comparison between the conventional full-storage time-domain adjoint method and the proposed Nyquist-based downsampled method, reducing field-storage memory from 1.62 GB to 0.02 GB with a per-slice storage of 80 KB. 
(f) Normalized focal-plane intensity profiles and (g) corresponding field-intensity maps at five representative wavelengths, demonstrating consistent broadband focusing performance.
}
  \label{fig:metalens2d}
\end{figure}

As a second benchmark, we optimized a broadband two-dimensional metalens using the proposed Nyquist-sampled time-domain adjoint method~\cite{Chung:20,yasuda2021design,wang2018broadband,park2025time}. The simulation domain is $5.2\,\mu\mathrm{m}$ wide and $4\,\mu\mathrm{m}$ tall, and the design region consists of a $5\,\mu\mathrm{m}\times0.2\,\mu\mathrm{m}$ thin layer with relative permittivity $\varepsilon_1=6$. A normally incident plane wave illuminates the metalens from below, and the objective maximizes the focusing efficiency for a numerical aperture of $\mathrm{NA}=0.6$ over the visible wavelength range of $500$--$700\,\mathrm{nm}$. Figure~\ref{fig:metalens2d}(b) summarizes the full set of simulation and design parameters.

The optimized layout in Fig.~\ref{fig:metalens2d}(a) forms a compact freeform dielectric structure that focuses normally incident broadband light. The focusing-efficiency spectrum in Fig.~\ref{fig:metalens2d}(c) compares the in-house FDTD solver with the open-source Meep FDTD package~\cite{oskooi2010meep}. The two solvers show close agreement across the target band. The optimized metalens achieves a focusing efficiency of approximately 55--60\% at shorter wavelengths and maintains broadband focusing performance with a gradual efficiency decrease toward $700\,\mathrm{nm}$, reflecting the broadband trade-offs imposed by the thin design layer and finite device aperture.

The storage comparison in Figs.~\ref{fig:metalens2d}(d) and~\ref{fig:metalens2d}(e) demonstrates the memory efficiency of the proposed method. The conventional full-storage implementation stores an 80~KB field snapshot at every FDTD time step, resulting in a total forward-field storage requirement of 1.62~GB. In contrast, the proposed method stores the same 80~KB snapshot only at Nyquist-sampled temporal intervals, reducing the required storage to 0.02~GB. This corresponds to an $81\times$ reduction in field storage memory while preserving the adjoint-gradient accuracy required for broadband optimization.

The focal-plane intensity profiles in Fig.~\ref{fig:metalens2d}(f) further verify the optical response at five representative wavelengths: $\lambda=500$, 550, 600, 650, and 700~nm. The in-house FDTD and Meep results produce nearly identical normalized intensity profiles, including the main focal peak and sidelobe structure. The corresponding two-dimensional field-intensity maps $|\mathbf{E}|^2$ in Fig.~\ref{fig:metalens2d}(g) confirm stable broadband focusing behavior throughout the visible band. These results demonstrate that the proposed memory-efficient adjoint framework can optimize broadband metalenses with substantial storage reduction while maintaining cross-solver consistency and high-quality focusing performance~\cite{Chung:20,wang2018broadband}.

\begin{figure}
  \centering
  \includegraphics[width=\textwidth]{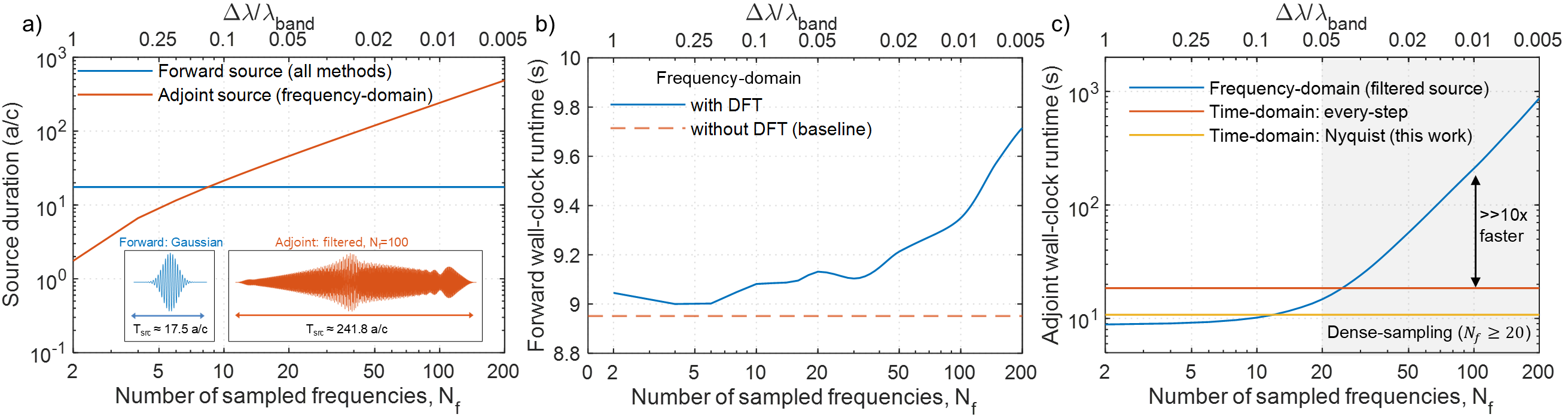}
  \caption{
  Runtime scaling for the broadband 2D metalens
  ($\mathrm{NA}=0.6$, $500$--$700\,\mathrm{nm}$).
  The bottom axis denotes the number of sampled frequencies $N_f$, and the
  top axis the normalized spectral interval
  $\Delta\lambda/\lambda_{\mathrm{band}}$, with
  $\lambda_{\mathrm{band}}=200\,\mathrm{nm}$.
  (a)~Source duration versus spectral sampling density.
  The forward Gaussian pulse has fixed duration
  ($T_{\mathrm{src}}\approx17.5\,a/c$), whereas the frequency-domain
  filtered adjoint source~\cite{hammond2022high} lengthens approximately as
  $1/\Delta f$; at $N_f=100$ it reaches $241.8\,a/c$, about $14\times$ the
  forward pulse. Insets: waveforms on a common time scale.
  (b)~Forward wall-clock runtime with and without DFT accumulation,
  showing less than $10\%$ variation ($8.95$--$9.72\,\mathrm{s}$) over the
  sampled range.
  (c)~Adjoint wall-clock runtime for the frequency-domain filtered-source
  method, conventional time-domain adjoint method~\cite{chung2000optimal},
  and proposed Nyquist-sampled method.
  The time-domain runtimes remain nearly independent of $N_f$, whereas the
  filtered-source runtime increases with spectral resolution; at $N_f=200$,
  the proposed method is approximately $80\times$ faster.
  All benchmarks use identical Meep FDTD settings.
  }
  \label{fig:runtime}
\end{figure}

The broadband 2D metalens provides a representative benchmark for identifying
the origin of the runtime difference between frequency-domain and time-domain
adjoint formulations as the spectral sampling density is increased.
Figure~\ref{fig:runtime} separates this comparison into three components:
the temporal duration of the forward and adjoint sources, the
frequency-dependent overhead in the forward simulation, and the resulting
adjoint-simulation runtime.
The bottom axis in each panel denotes the number of sampled frequencies
$N_f$, while the top axis denotes the corresponding normalized spectral
interval $\Delta\lambda/\lambda_{\mathrm{band}}$, where
$\lambda_{\mathrm{band}}=\lambda_{\max}-\lambda_{\min}$.
For the present $500$--$700\,\mathrm{nm}$ target band,
$\lambda_{\mathrm{band}}=200\,\mathrm{nm}$, such that increasing $N_f$
corresponds to progressively finer spectral sampling.

Figure~\ref{fig:runtime}(a) first compares the temporal durations of the
sources used by the different formulations.
The same broadband Gaussian pulse is used for the forward simulation in all
methods, and its duration remains fixed at approximately
$T_{\mathrm{src}}=17.5\,a/c$, independent of $N_f$.
The frequency-domain filtered-source formulation, by contrast, synthesizes
an adjoint excitation whose complex spectral amplitudes are prescribed at
the sampled frequencies~\cite{hammond2022high}.
As the frequency spacing $\Delta f$ is reduced, the corresponding temporal
waveform must extend over an increasingly long interval, with a characteristic
duration scaling as $T_{\mathrm{src}}\sim1/\Delta f$.
This trend is directly visible in Fig.~\ref{fig:runtime}(a).
For example, at $N_f=100$, the filtered adjoint source has a duration of
approximately $241.8\,a/c$, compared with $17.5\,a/c$ for the forward
Gaussian pulse, corresponding to an approximately $14\times$ increase in
source duration.
The representative waveforms shown in the insets illustrate this temporal
extension on a common time scale.

The forward-runtime comparison in Fig.~\ref{fig:runtime}(b) isolates the
additional cost associated with frequency-domain DFT accumulation.
When DFT accumulation is removed, the underlying FDTD propagation time
remains essentially independent of $N_f$.
Including the DFT monitors introduces a gradual increase in wall-clock time
because the field amplitudes must be accumulated at an increasing number of
sampled frequencies.
However, the total variation remains below $10\%$ over the range
$2\leq N_f\leq200$, from approximately $8.95$ to $9.72\,\mathrm{s}$.
Thus, although DFT accumulation introduces a measurable
sampling-dependent overhead, the forward pass alone does not account for the
large runtime separation observed at dense spectral sampling.

The dominant difference instead appears during the adjoint simulation, as
shown in Fig.~\ref{fig:runtime}(c).
For the frequency-domain filtered-source formulation, the adjoint wall-clock
time increases rapidly with $N_f$, consistent with the increasing temporal
extent of the synthesized adjoint source shown in
Fig.~\ref{fig:runtime}(a).
In contrast, both time-domain formulations evaluate the broadband gradient
using a single broadband adjoint simulation~\cite{chung2000optimal,nomura2007structural}.
Their adjoint simulation duration is therefore determined by the underlying
time-domain problem rather than by the number of discrete spectral samples,
and their runtimes remain nearly constant as $N_f$ is increased.

At coarse spectral sampling, the frequency-domain filtered-source formulation
remains computationally competitive because the synthesized adjoint source is
relatively short.
As the sampling becomes denser, however, the source duration and corresponding
adjoint runtime increase rapidly.
For the present comparison, the shaded region in
Fig.~\ref{fig:runtime}(c) denotes $N_f\geq20$, corresponding to
$\Delta\lambda/\lambda_{\mathrm{band}}\leq0.05$ and representative of the
dense spectral sampling commonly used to characterize broadband device
responses~\cite{Chung:20}.
Within this regime, the runtime separation between the frequency-domain and
time-domain formulations increases continuously with spectral resolution.
At $N_f=200$, the proposed Nyquist-sampled time-domain method requires
approximately $80\times$ less adjoint wall-clock time than the
frequency-domain filtered-source formulation.

Taken together, Figs.~\ref{fig:runtime}(a)--\ref{fig:runtime}(c) show that
the spectral-sampling-dependent runtime cost of the filtered-source
frequency-domain implementation arises primarily in the adjoint stage.
Forward DFT accumulation introduces only a modest overhead, whereas the
filtered-source construction incurs a progressively longer temporal support;
the Supporting Information further shows an $N_f$-dependent increase in the
per-step cost of evaluating the multi-basis source.
The time-domain formulations avoid this spectral-sampling-dependent adjoint
cost because the broadband response is evaluated directly from a single
forward--adjoint simulation pair.
The proposed Nyquist-sampled formulation retains this runtime advantage while
simultaneously reducing the field-storage burden, as quantified in
Table~\ref{tab:2d_memory}; extended three-dimensional scaling is provided in
Section~S1 (Fig.~S1) of the Supporting Information.

% == 2D Wavelength demultiplexer ==
\subsubsection{2D wavelength demultiplexer}
\label{subsubsec:demux}

\begin{figure}
  \centering
  \includegraphics[width=\textwidth]{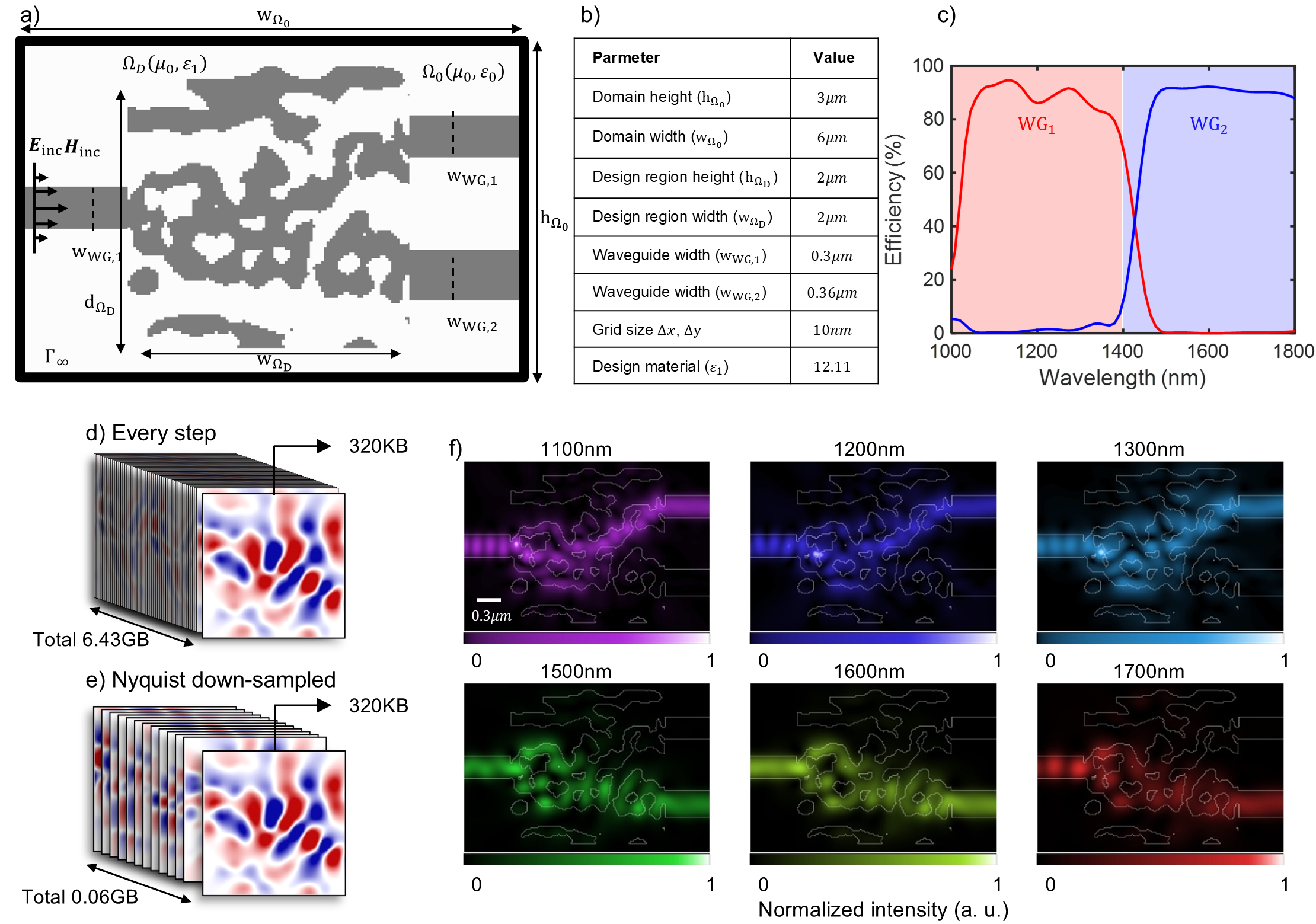}
\caption{
Broadband wavelength-demultiplexer demonstration. 
(a) Schematic of the optimized two-dimensional demultiplexer with one input waveguide and two output waveguides. 
(b) Key simulation and design parameters. 
(c) Transmission spectra into WG$_1$ and WG$_2$, demonstrating efficient routing of short- and long-wavelength bands to different output ports. 
(d,e) Memory comparison between the conventional full-storage time-domain adjoint method and the proposed Nyquist-based downsampled method, reducing field-storage memory from 6.43 GB to 0.06 GB with a per-slice storage of 320 KB. 
(f) Normalized field-intensity distributions at six representative wavelengths, showing wavelength-selective routing to WG$_1$ for 1100--1300 nm and to WG$_2$ for 1500--1700 nm.
}
 
  \label{fig:demux}
\end{figure}

As a third benchmark, we optimized a broadband two-dimensional wavelength demultiplexer using the proposed Nyquist-sampled time-domain adjoint method~\cite{piggott2015inverse,su2018inverse}. The simulation domain spans $6\,\mu\mathrm{m}\times3\,\mu\mathrm{m}$ and contains a $2\,\mu\mathrm{m}\times2\,\mu\mathrm{m}$ freeform silicon design region with relative permittivity $\varepsilon_1=12.11$ in an air background. A broadband signal enters through a $0.3\,\mu\mathrm{m}$-wide input waveguide, and the optimization objective simultaneously maximizes wavelength-selective transmission into two output waveguides. The first output waveguide, WG$_1$, with width $w_{\mathrm{WG},1}=0.3\,\mu\mathrm{m}$, targets the short-wavelength band of approximately $1000$--$1300\,\mathrm{nm}$, whereas the second output waveguide, WG$_2$, with width $w_{\mathrm{WG},2}=0.36\,\mu\mathrm{m}$, targets the long-wavelength band of approximately $1400$--$2000\,\mathrm{nm}$. Figure~\ref{fig:demux}(b) summarizes the full set of simulation and design parameters.

The optimized geometry in Fig.~\ref{fig:demux}(a) forms a compact freeform silicon structure that separates the input broadband signal into the two designated output channels. The wavelength-dependent routing efficiencies in Fig.~\ref{fig:demux}(c) show that the device achieves near-unity transmission in each target band while maintaining clear channel selectivity. These results confirm that the proposed memory-efficient adjoint method accurately captures the broadband spectral-routing objective and guides the optimization toward a high-performance demultiplexing structure.

The storage comparison in Figs.~\ref{fig:demux}(d) and~\ref{fig:demux}(e) demonstrates the memory reduction achieved by the proposed method. The conventional full-storage implementation stores a 320~KB field snapshot at each FDTD time step, resulting in a total forward-field storage requirement of 6.43~GB. In contrast, the proposed method stores the same 320~KB snapshot only at Nyquist-sampled temporal intervals, reducing the required storage to 0.06~GB. This corresponds to an approximately $107\times$ reduction in field storage memory while preserving the adjoint-gradient accuracy required for broadband wavelength-selective optimization.

The normalized field-intensity maps $|\mathbf{E}|^2$ in Fig.~\ref{fig:demux}(f) further verify the wavelength-dependent routing behavior at six representative wavelengths: 1100, 1200, 1300, 1500, 1600, and 1700~nm. At short wavelengths, the field concentrates toward WG$_1$, whereas at long wavelengths, the field redirects toward WG$_2$. The clear spatial separation of the output channels across the operating band demonstrates that the proposed Nyquist-sampled adjoint framework can optimize compact broadband wavelength demultiplexers while reducing the field-storage burden by more than two orders of magnitude.

% == 2D Color router ==
\subsubsection{2D color router}
\label{subsubsec:colorrouter}

\begin{figure}
  \centering
  \includegraphics[width=\textwidth]{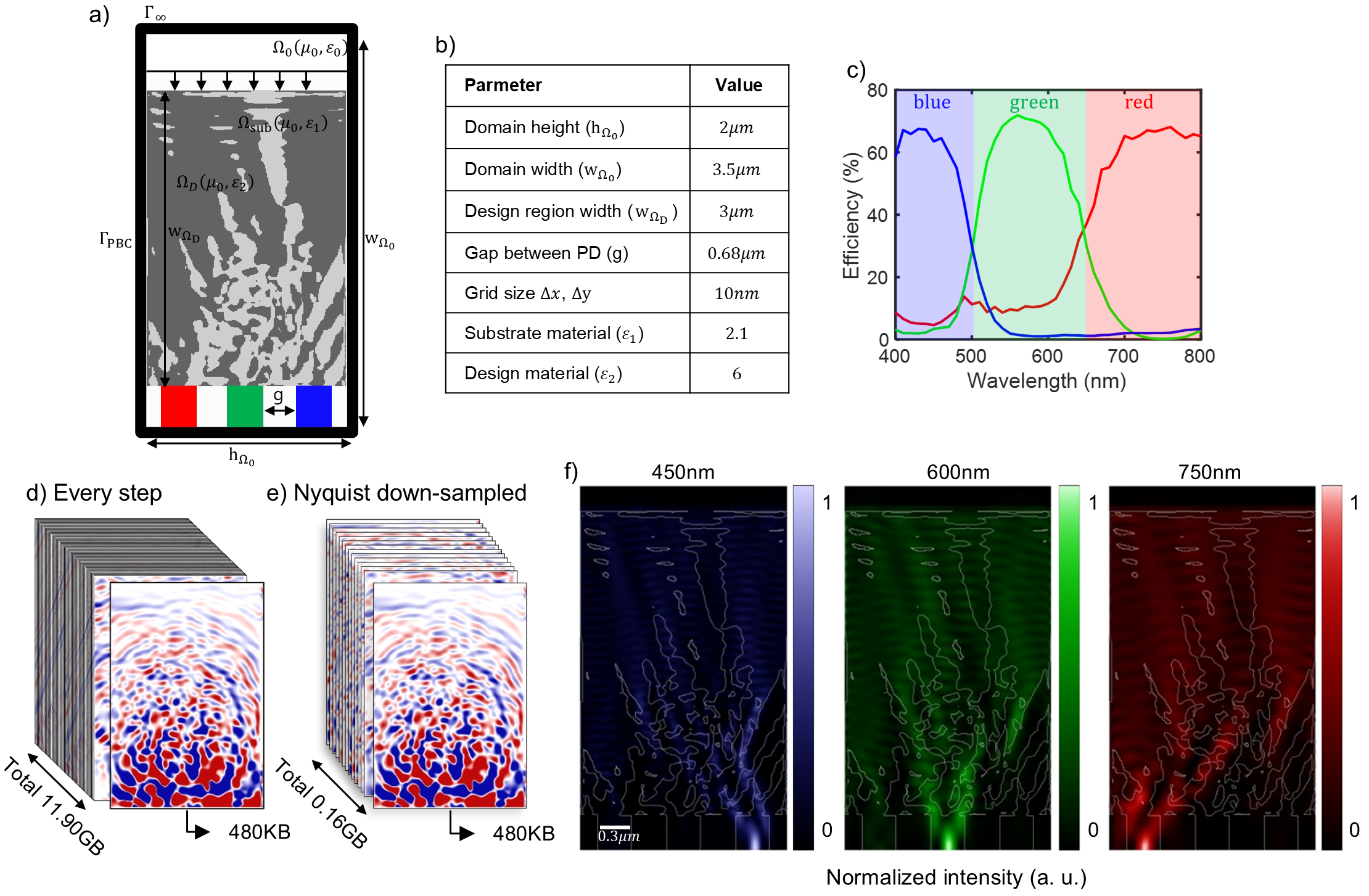}
  \caption{
Visible-band color-router demonstration. 
(a) Schematic of the optimized two-dimensional color router with a freeform design region on a substrate and three bottom pixel detectors for blue, green, and red channels. 
(b) Key simulation and design parameters. 
(c) Coupling efficiency spectra into the three output channels over 400--800 nm, demonstrating wavelength-selective routing. 
(d,e) Memory comparison between the conventional full-storage time-domain adjoint method and the proposed Nyquist-based downsampled method, reducing field-storage memory from 11.90 GB to 0.16 GB with a per-slice storage of 480 KB. 
(f) Normalized field-intensity distributions at 450, 600, and 750 nm, confirming selective routing to the blue, green, and red output channels, respectively.
}
 
  \label{fig:colorrouter}
\end{figure}

As a fourth benchmark, we optimized a two-dimensional color router for spectrally selective spatial routing across the visible wavelength range of approximately $400$--$800\,\mathrm{nm}$~\cite{kim2024freeform,lee2024inverse}. The simulation uses periodic boundary conditions $\Gamma_{\mathrm{PBC}}$ along the lateral direction, with a domain width of $3.5\,\mu\mathrm{m}$ and a height of $2\,\mu\mathrm{m}$. The device consists of a freeform design layer with relative permittivity $\varepsilon_2=6$ placed above a substrate region $\Omega_{\mathrm{sub}}$ with relative permittivity $\varepsilon_1=2.1$. A normally incident plane wave illuminates the structure from above. The optimization objective simultaneously maximizes the coupling efficiency into three spatially separated pixel detectors at the bottom plane, corresponding to blue, green, and red target channels centered near 450, 600, and 750~nm, respectively. Adjacent detectors are separated by a gap of $g=0.68\,\mu\mathrm{m}$. Figure~\ref{fig:colorrouter}(b) summarizes the full set of simulation and design parameters.

The optimized geometry in Fig.~\ref{fig:colorrouter}(a) forms a compact freeform dielectric structure that redirects incident light toward different detector positions depending on wavelength. The wavelength-dependent coupling efficiencies in Fig.~\ref{fig:colorrouter}(c) show clearly separated spectral responses for the blue, green, and red output channels. Each channel reaches a peak routing efficiency exceeding 60\%, demonstrating that the proposed Nyquist-sampled adjoint method successfully optimizes a multi-channel visible-band routing objective with strong wavelength selectivity.

The storage comparison in Figs.~\ref{fig:colorrouter}(d) and~\ref{fig:colorrouter}(e) further demonstrates the memory advantage of the proposed framework. In the conventional full-storage implementation, each field snapshot requires 480~KB, and storing the forward-field history at every FDTD time step results in a total storage requirement of 11.90~GB. This case represents the largest two-dimensional memory footprint among the benchmarks considered in this study. In contrast, the proposed method stores the same 480~KB snapshot only at Nyquist-sampled temporal intervals, reducing the total storage to 0.16~GB. This corresponds to a $74\times$ reduction in field storage memory while retaining the adjoint-gradient accuracy needed for broadband color-routing optimization.

The normalized field-intensity maps $|\mathbf{E}|^2$ in Fig.~\ref{fig:colorrouter}(f) verify the wavelength-selective spatial routing behavior at three representative wavelengths: 450, 600, and 750~nm. At 450~nm, the field concentrates toward the left detector corresponding to the blue channel. At 600~nm, the field is directed toward the middle detector, corresponding to the green channel. At 750~nm, the field couples preferentially into the right detector corresponding to the red channel. These results demonstrate that the proposed memory-efficient adjoint framework can optimize compact visible-band color routers with multi-channel spectral selectivity while reducing the field-storage burden by nearly two orders of magnitude~\cite{kim2024freeform,lee2024inverse}.

% == 3D Metalens ==
\subsubsection{3D broadband metalens}
\label{subsubsec:metalens3d}

\begin{figure}
  \centering
  \includegraphics[width=\textwidth]{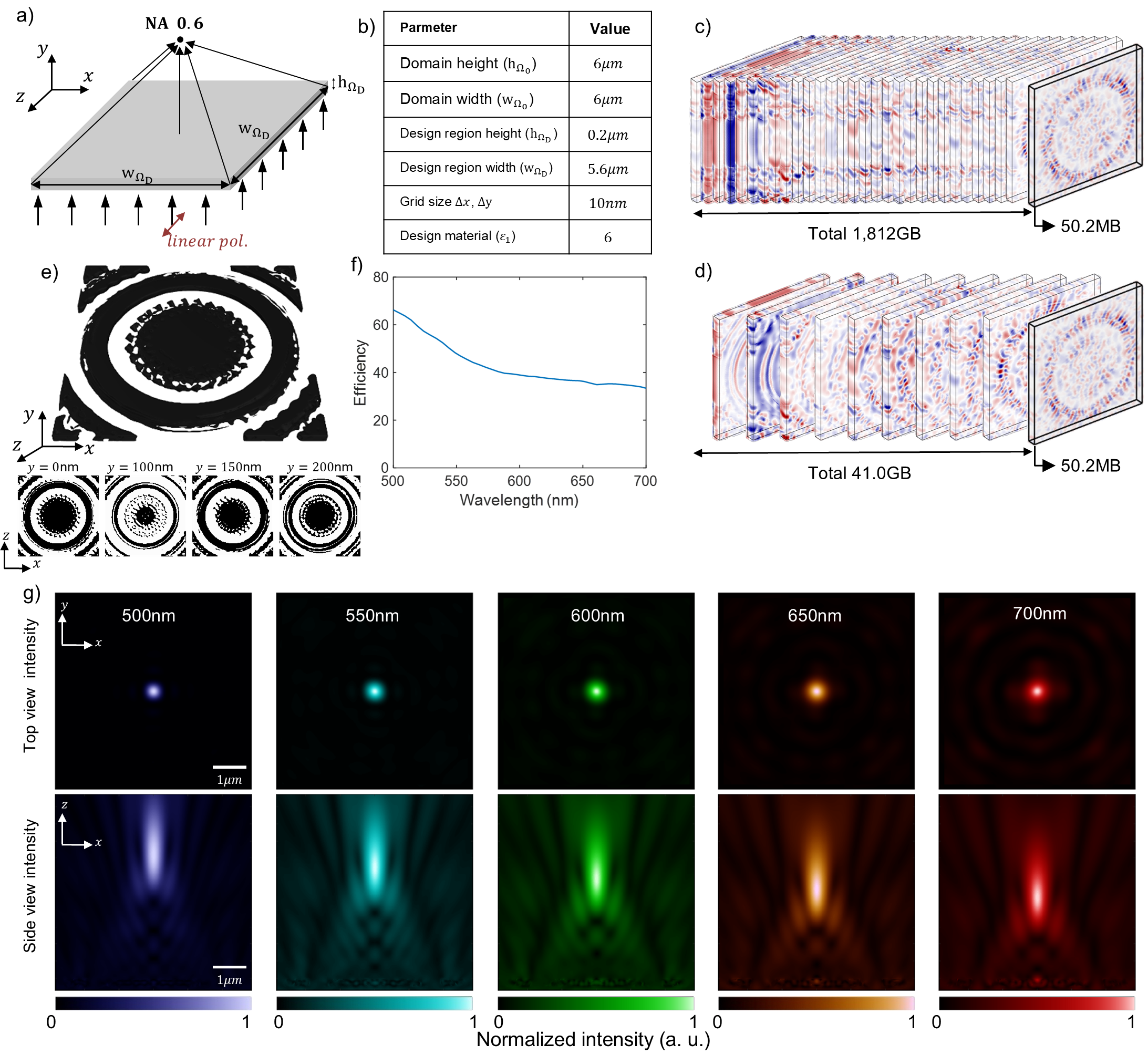}
  \caption{
Three-dimensional broadband metalens demonstration. 
(a) Schematic of the 3D metalens with $\mathrm{NA}=0.6$ under normally incident linearly polarized illumination. 
(b) Key simulation and design parameters. 
(c,d) Memory comparison between the conventional full-storage time-domain adjoint method and the proposed Nyquist-based downsampled method. By storing only Nyquist-rate forward-field samples, the proposed method reduces the total field-storage requirement from 1,812 GB to 41.0 GB, while each field slice occupies 50.2 MB. 
(e) Optimized 3D metalens topology, including a top view and cross-sectional slices at $y=0$, 100, 150, and 200 nm. 
(f) Focusing efficiency spectrum over 500--700 nm. 
(g) Normalized focal-plane and longitudinal intensity distributions at five representative wavelengths, demonstrating broadband on-axis focusing.
}

  \label{fig:metalens3d}
\end{figure}

As a final large-scale demonstration, we apply the proposed Nyquist-sampled time-domain adjoint method to the inverse design of a fully three-dimensional broadband metalens with $\mathrm{NA}=0.6$~\cite{hughes2021perspective,phan2019high,mansouree2021large,sang2022toward,Chung:20}. The simulation domain spans $6\,\mu\mathrm{m}\times6\,\mu\mathrm{m}\times6\,\mu\mathrm{m}$ and contains a thin design layer with lateral width $w_{\Omega_D}=5.6\,\mu\mathrm{m}$ and thickness $h_{\Omega_D}=0.2\,\mu\mathrm{m}$. We pattern the design layer from a dielectric material with relative permittivity $\varepsilon_1=6$ on a 10~nm grid, yielding approximately $6.97\times10^6$ design voxels. A normally incident linearly polarized plane wave illuminates the metalens from below, as illustrated in Fig.~\ref{fig:metalens3d}(a), and Fig.~\ref{fig:metalens3d}(b) summarizes the full set of simulation and design parameters. 
The extended scaling analysis in Section~S1 (Fig.~S1) of the Supporting Information places this problem size beyond typical single-node memory capacities for conventional full-history storage~\cite{kang2024large}. The proposed method therefore enables a class of large-scale three-dimensional broadband inverse-design problems that would otherwise require prohibitive field-storage memory.

The storage comparison in Figs.~\ref{fig:metalens3d}(c) and~\ref{fig:metalens3d}(d) shows the memory reduction achieved by the proposed method. In the conventional full-storage implementation, each field snapshot requires 50.2~MB, and storing the forward-field history at every FDTD time step accumulates to 1{,}812~GB. This requirement far exceeds the memory capacity of high-end GPUs and typical large-memory CPU nodes. In contrast, the proposed method stores the same 50.2~MB snapshot only at Nyquist-sampled temporal intervals, reducing the total storage to 41.0~GB. This corresponds to an approximately $44\times$ reduction in forward-field memory while preserving the adjoint-gradient accuracy required for broadband optimization.

The optimized three-dimensional topology in Fig.~\ref{fig:metalens3d}(e) includes a top view of the full aperture and cross-sectional slices through the design layer at $y=0$, 100, 150, and 200~nm. The resulting structure exhibits a complex concentric pattern reminiscent of Fresnel-zone behavior~\cite{pan2022dielectric}, together with fine-scale features distributed across the aperture. This geometry shows that the proposed method can recover physically meaningful broadband focusing structures at a multi-million-voxel three-dimensional scale. The focusing-efficiency spectrum in Fig.~\ref{fig:metalens3d}(f) decreases gradually from approximately 65\% at 500~nm to approximately 35\% at 700~nm under $x$-polarized illumination, demonstrating broadband visible focusing within the optimized NA.

The normalized intensity distributions in Fig.~\ref{fig:metalens3d}(g) further verify the broadband focusing behavior at $\lambda=500$, 550, 600, 650, and 700~nm. The focal-plane $xy$ intensity profiles in the top row show on-axis focusing at each wavelength, while the longitudinal $xz$ intensity profiles in the bottom row show the corresponding propagation and focal profiles. The focal spot remains well defined at shorter wavelengths and gradually broadens with reduced contrast at longer wavelengths, consistent with the efficiency trend in Fig.~\ref{fig:metalens3d}(f).

This three-dimensional demonstration confirms the practical applicability of the proposed Nyquist-sampled time-domain adjoint method to large-scale 3D broadband metalens inverse design~\cite{phan2019high,sang2022toward}. In a regime where the conventional method requires terabyte-scale field storage, the proposed method preserves broadband adjoint optimization capability while reducing the memory demand to a tractable level for modern high-memory GPU systems.

% ============================================================
\section{Discussion and Outlook}
\label{sec:conclusion}

We have presented a Nyquist-sampled time-domain adjoint optimization framework for memory-efficient broadband nanophotonic inverse design. The proposed method is based on the observation that FDTD field histories generated by band-limited excitations are temporally oversampled relative to the Nyquist rate required for gradient recovery. By storing the forward field only at Nyquist-compliant temporal intervals and accumulating adjoint-gradient contributions on the fly during the adjoint simulation, the method substantially reduces the forward-field storage requirement while eliminating the need to store the adjoint-field history, preserving the \(1+1\) simulation count of conventional time-domain adjoint optimization.

Numerical verification confirms that the proposed downsampling strategy preserves adjoint-gradient accuracy when the sampling interval satisfies the Nyquist limit, whereas aliasing-induced errors emerge when this limit is exceeded. Across representative two-dimensional devices, including waveguide bends, metalenses, wavelength demultiplexers, and color routers, the method reduces memory consumption by one to two orders of magnitude without degrading optimized device performance. A fully three-dimensional broadband metalens further demonstrates the practical scalability of the approach, reducing the required field-storage memory from terabyte-scale levels to a range accessible with modern high-memory computing hardware.

Beyond memory efficiency, the runtime comparison further clarifies where the proposed framework is most advantageous. Both time-domain methods evaluate the entire target band using a single forward--adjoint simulation pair, making their runtimes essentially independent of \(N_f\). By contrast, the synthesized adjoint-source duration in the frequency-domain filtered-source approach scales as \(1/\Delta f_{\min}\), or approximately linearly with \(N_f\) for a fixed bandwidth. The total adjoint runtime does not necessarily follow the same scaling, as it also depends on the device-dependent field-decay time and the stopping criterion. This constant-time broadband scaling makes time-domain adjoint optimization increasingly favorable as the required spectral density grows, and the proposed method retains this advantage while further lowering the per-iteration runtime by reducing field-storage overhead. The two contributions are therefore complementary: Nyquist sampling reduces the memory footprint that has historically restricted time-domain adjoint optimization, while the underlying time-domain formulation simultaneously preserves the runtime efficiency that distinguishes it from frequency-domain approaches for densely sampled broadband objectives.

Taken together, these results show that the dominant memory bottleneck in conventional time-domain adjoint optimization is not an intrinsic requirement of broadband gradient evaluation, but rather a consequence of redundant temporal field storage. The proposed Nyquist-sampled formulation, therefore, provides a simple and general route toward scalable broadband inverse design that is efficient in both memory and runtime. The present implementation evaluates the gradient by direct quadrature on the downsampled temporal grid; alternative evaluation schemes---band-limited (sinc) interpolation back to the full FDTD grid, higher-order quadrature rules, or adaptive per-region sampling rates---may further improve accuracy near the Nyquist boundary and merit systematic investigation. Beyond nanophotonics, the same principle may be applicable to other time-domain adjoint problems in electromagnetics and wave physics, provided that the relevant field histories are band-limited or can be reliably characterized in frequency.

% outlook
Large-scale full-wave simulations of Maxwell's equations have historically been driven by applications in computational electromagnetics, including radar scattering~\cite{taflove1975numerical,song1997multilevel}, antenna design~\cite{harrington1993field,jin2015finite}, aerospace systems~\cite{volakis1998finite,song1997multilevel}, and defense-related technologies~\cite{chew2001fast,sankaran2019recent}, where both the computational cost and practical importance are exceptionally high. 
Similar computational challenges are now becoming important to optics and photonics, as inverse-designed optical components are increasingly viewed as potential alternatives to conventional industrial solutions. For example, large-area metalenses and metaholograms are emerging as promising platforms for compact imaging, display, and wavefront-control systems, making it increasingly important to optimize metasurface efficiency toward fundamental physical limits over device footprints spanning many wavelengths. Photonic integrated circuits are likewise being developed as a route to overcome the bandwidth, latency, and energy-efficiency limitations of electrical interconnects in data-intensive computing systems, while photonic quantum technologies offer a promising platform for scalable quantum information processing. Across these applications, the relevant design spaces are becoming too large to be explored by intuition-driven design or conventional optimization alone, and the broadband objectives they entail often demand dense spectral sampling, where the constant-time scaling of time-domain adjoint optimization becomes particularly valuable. Advanced adjoint-based optimization methods, combined with multi-node high-performance computing systems equipped with high-bandwidth memory and distributed-memory architectures, may therefore provide an essential foundation for discovering large-scale, high-efficiency optical and photonic devices. In this context, algorithmic strategies that jointly reduce redundant data storage, memory movement, and per-iteration computational cost, such as the Nyquist-sampled time-domain adjoint framework proposed here, are expected to play an increasingly important role in making practical large-scale photonic inverse design computationally tractable.

\section*{Supplementary Information}

Additional benchmark and scaling data are provided in the Supplementary
Information accompanying this preprint. The supplementary document contains
extended two- and three-dimensional memory-scaling analysis and a controlled
Meep runtime-mechanism validation of the frequency- and time-domain adjoint
benchmarks, including source-duration, FDTD-step, per-step-cost, and
DFT-monitor controls.

\bibliographystyle{apsrev4-2}
\bibliography{reference}

\end{document}

\maketitle

This Supporting Information contains (i) an extended memory-scaling analysis
for two- and three-dimensional inverse-design problems and (ii) a controlled
Meep benchmark~\cite{oskooi2010meep} that decomposes the runtime scaling of
frequency-domain filtered-source and time-domain adjoint formulations.

\section{Memory Scaling of Nyquist-Sampled Adjoint Optimization}
\label{sec:memory}

This section quantifies the field-storage memory of the proposed
Nyquist-sampled time-domain adjoint method relative to the conventional
full-storage time-domain adjoint method, across representative
two-dimensional benchmark devices and projected three-dimensional
inverse-design problems. Whereas Figure~1 of the main text illustrates the
conceptual differences among the adjoint formulations, the comparison here
evaluates their practical implications for computational feasibility, showing
how Nyquist-sampled forward-field storage and on-the-fly gradient
accumulation reduce the dominant memory requirement as the number of design
cells or voxels increases. These comparisons quantify the storage cost of the time-domain formulations
discussed in the main text. The frequency-domain storage baseline can be
smaller when only a few frequencies are sampled, but the corresponding
runtime cost increases as the prescribed spectral sampling is refined, as
examined in the Meep runtime benchmark in
Section~\ref{sec:runtime-validation}.

% The runtime-validation figure has been removed; this is the only SI figure
% and is therefore numbered Fig. S1.
\begin{figure}[!htbp]
\centering
\includegraphics[width=\textwidth]{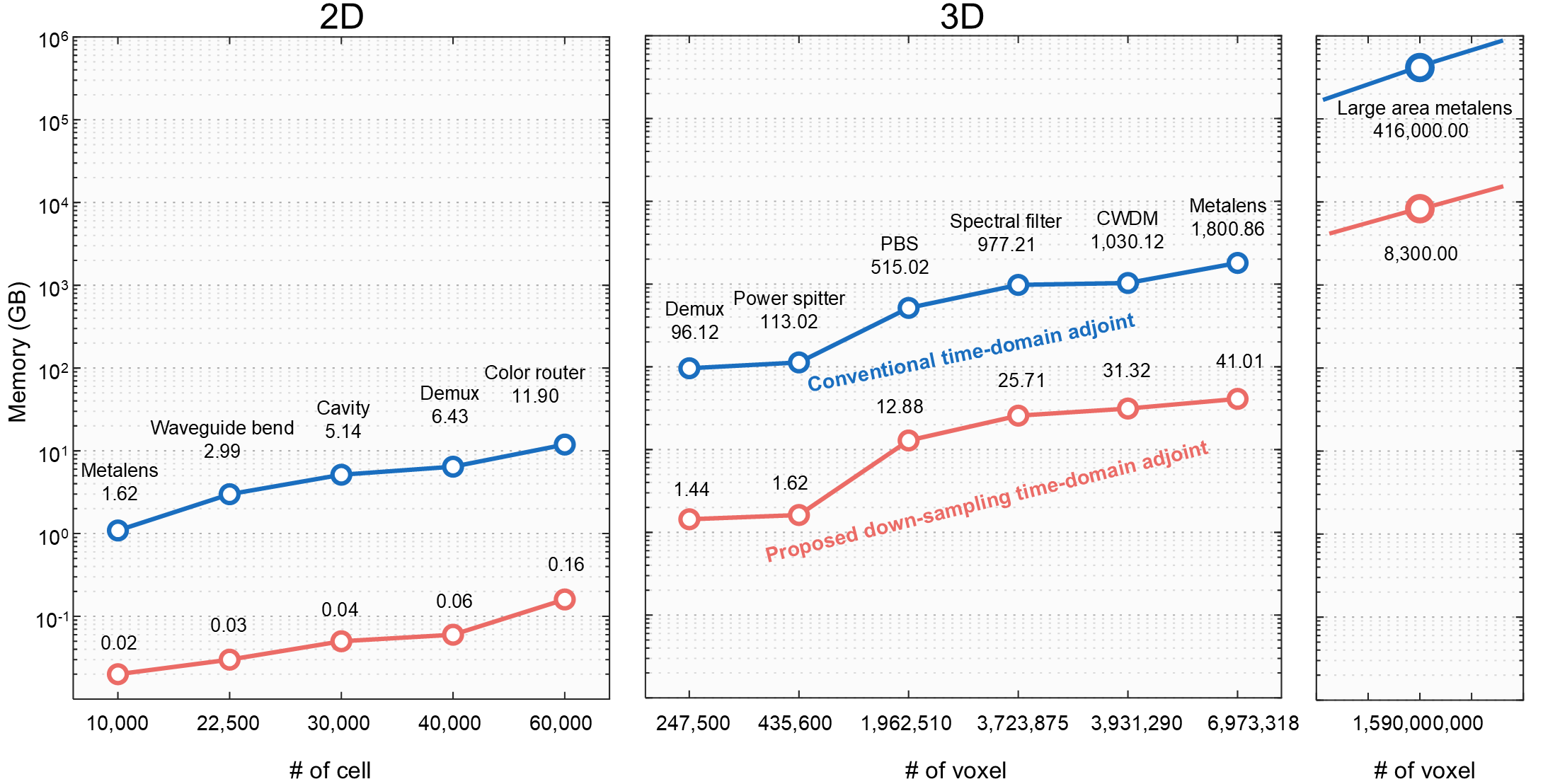}%
\caption{Memory-scaling comparison between the conventional time-domain
adjoint method and the proposed Nyquist-sampled time-domain adjoint method.
The left panel shows two-dimensional benchmark devices, while the middle
panel shows the predicted memory requirements for representative
three-dimensional nanophotonic devices as the number of design voxels
increases. The right panel extrapolates the memory demand to a large-area
metalens. Blue and red curves denote the conventional time-domain adjoint
method and the proposed method, respectively. The proposed method
substantially reduces the dominant field-storage memory requirement while
preserving the two-simulation advantage of time-domain adjoint optimization,
enabling large-scale three-dimensional inverse-design problems that are
otherwise inaccessible with full time-step field storage.}
\label{fig:memory}
\end{figure}

The left panel of Figure~\ref{fig:memory} reports the measured memory usage
for representative two-dimensional benchmark devices, whereas the center and
right panels extend the same scaling analysis to three-dimensional
inverse-design problems using device sizes reconstructed from previously
reported photonic inverse-design
studies~\cite{piggott2015inverse,su2018inverse,phan2019high,mansouree2021large,hughes2021perspective}.
In all cases, both methods exhibit memory requirements that scale
approximately linearly with the number of design cells or voxels. The key
difference lies in the proportionality constant: Nyquist-sampled
forward-field storage and on-the-fly gradient accumulation substantially
reduce the amount of field data that must be retained during the adjoint
calculation.

\subsection{Two-dimensional benchmarks}

For the two-dimensional benchmarks, the conventional time-domain adjoint
method requires 1.62~GB for the metalens with 10{,}000 design cells,
2.99~GB for the waveguide bend with 22{,}500 cells, 5.14~GB for the cavity
with 30{,}000 cells, 6.43~GB for the demultiplexer with 40{,}000 cells, and
11.90~GB for the color router with 60{,}000 cells. In contrast, the proposed
method reduces these memory requirements to 0.02~GB, 0.03~GB, 0.04~GB,
0.06~GB, and 0.16~GB, respectively. These results demonstrate a nearly
two-order-of-magnitude memory reduction, indicating that the full time-step
field history used in the conventional implementation contains substantial
temporal redundancy for bandwidth-limited responses. By storing only the
forward-field samples required to satisfy the Nyquist
criterion~\cite{shannon1949communication} and accumulating the gradient
during the adjoint simulation, the proposed method keeps all two-dimensional
benchmarks below 0.2~GB while preserving the two-simulation structure of the
time-domain adjoint formulation.

\subsection{Projected three-dimensional problems}

The advantage becomes more pronounced in three dimensions, where the
conventional full-storage approach rapidly reaches memory requirements beyond
typical single-GPU and single-node CPU capacities. As shown in the center
panel of Figure~\ref{fig:memory}, the projected memory requirement of the
conventional method increases from 96.12~GB for a 3D demultiplexer with
247{,}500 voxels and 113.02~GB for a 3D power splitter with 435{,}600 voxels,
both already exceeding a representative 80~GB GPU memory capacity, to
515.02~GB for a polarization beam splitter with 1{,}962{,}510 voxels. The
requirement further rises to 977.21~GB for a spectral filter with
3{,}723{,}875 voxels, 1{,}030.12~GB for a coarse wavelength-division
multiplexer with 3{,}931{,}290 voxels, and 1{,}800.86~GB for a 3D metalens
with 6{,}973{,}318 voxels. These values approach or exceed a representative
1~TB CPU memory capacity, making conventional full-history storage impractical
for large three-dimensional inverse-design problems. For the large-area
metalens shown in the right panel, which contains $1.59\times10^{9}$ voxels,
the conventional method would require approximately 416{,}000~GB, or 416~TB,
placing it far beyond the capacity of typical single-node CPU or GPU systems.

In contrast, the proposed Nyquist-sampled time-domain adjoint method
substantially reduces the projected memory requirements for the same devices.
The required memory decreases to 1.44~GB for the demultiplexer, 1.62~GB for
the power splitter, 12.88~GB for the polarization beam splitter, 25.71~GB for
the spectral filter, 31.32~GB for the CWDM device, and 41.01~GB for the 3D
metalens. The large-area metalens case also decreases from 416~TB to
approximately 8.3~TB. Although this large-area example remains demanding and
would require distributed-memory or out-of-core implementations, the proposed
method reduces the memory requirement by roughly two orders of magnitude
compared with the conventional full-storage approach.

These results demonstrate the key practical advantage of the proposed
framework. By combining Nyquist-sampled forward-field storage with
on-the-fly gradient accumulation, the method preserves the two-simulation
efficiency of time-domain adjoint optimization while substantially reducing
the dominant field-storage burden. Several representative 3D devices,
including the demultiplexer, power splitter, polarization beam splitter,
spectral filter, CWDM device, and 3D metalens, move from a GPU-infeasible or
CPU-memory-limited regime to a memory range compatible with modern
high-memory GPUs. Thus, the proposed method expands the practical design
space for broadband three-dimensional photonic inverse design without
modifying the underlying FDTD update equations.

\section{Meep Runtime Validation of Frequency- and Time-Domain Adjoint Scaling}
\label{sec:runtime-validation}

\subsection{Scope and timing definition}

This Supporting Information section documents the Meep benchmark used to
separate the mechanisms responsible for the runtime scaling of the
frequency-domain filtered-source adjoint calculation~\cite{hammond2022high}
and to compare it with the conventional and Nyquist-sampled time-domain
adjoint implementations. The benchmark was
designed to answer three specific questions: (i) whether the temporal support
of the synthesized frequency-domain adjoint source increases as the number of
sampled frequencies, $\Nf$, is increased; (ii) whether the corresponding
wall-clock increase arises from a larger number of FDTD updates,
from a larger cost per update, or from running-DFT accumulation; and (iii)
whether the time-domain adjoint runtime remains independent of $\Nf$ when the
same physical device and FDTD discretization are used. The complementary
field-storage memory scaling of the proposed method is documented in
Section~\ref{sec:memory}.

All reported runtimes correspond to the wall-clock time spent in the FDTD
solve itself. For the frequency-domain calculations, incident-field
normalization, the untimed forward prerequisite used to construct the adjoint
source, adjoint-source construction, and post-processing were excluded from
the reported \texttt{Simulation.run} time. DFT updates performed during the
FDTD solve were included. For the time-domain calculations, the timed forward
and adjoint solves include the field-sampling and gradient-accumulation
callbacks executed during the simulation, whereas final post-run gradient
assembly is excluded. Thus, the comparison isolates the electromagnetic solve
and the per-time-step operations required by each adjoint formulation rather
than the complete optimization-loop overhead.

\subsection{Common physical and numerical configuration}

The benchmark uses the same two-dimensional $E_z$-polarized metalens geometry
for the frequency-domain (FD) and time-domain (TD) calculations. The physical
simulation region is $5.2a\times4.0a$, surrounded by a perfectly matched layer
(PML) of thickness $0.7a$. The design region is $5.0a\times0.2a$ and is
parameterized by a $501\times15$ material grid interpolating between
$\varepsilon=1$ and $\varepsilon=6$. A line source located at $y=-1.85a$
illuminates the structure, and the focal-point monitor is located $3.33a$ above
the top of the design layer. The random initial design is generated with a
fixed seed and mirror symmetry so that every benchmark condition uses the same
permittivity distribution.

The spatial resolution is 100 pixels/$a$ and the Courant factor is 0.5,
giving
\begin{equation}
\dtfdtd = 0.005\,a/c.
\end{equation}
The broadband forward excitation is a single Gaussian $E_z$ source centered at
$f_c=1.714286\,c/a$ with bandwidth parameter
$f_{\mathrm{width}}=0.571429\,c/a$ and cutoff 5. Its temporal support is
$17.5\,a/c$. The sampled wavelength interval is $0.50a$--$0.70a$
(500--700 nm when $a=1~\mu$m). For a specified $\Nf$, wavelengths are sampled
uniformly and converted to frequencies according to
\begin{equation}
\lambda_i = \lambda_{\min} +
\frac{i}{\Nf-1}(\lambda_{\max}-\lambda_{\min}),
\qquad
f_i = \frac{1}{\lambda_i},
\qquad i=0,\ldots,\Nf-1.
\end{equation}
The benchmark values are
$\Nf=2,5,10,15,20,30,50,100,150,$ and 200.

\begin{table}[htbp]
\caption{Common physical, numerical, termination, and execution settings for
the Meep benchmark.}
\label{tab:config}
\centering
\small
\begin{tabularx}{\textwidth}{>{\raggedright\arraybackslash}p{0.28\textwidth}X}
\toprule
Item & Setting \\
\midrule
Meep / PyMeep & Meep module 1.30.0; PyMeep 1.30.1, build \texttt{nompi\_py39h9d793ab\_103} \\
Sample counts & $\Nf=2,5,10,15,20,30,50,100,150,200$ \\
Wavelength / frequency band & $0.50$--$0.70a$ (500--700 nm); $1.428571$--$2.000000\,c/a$ \\
Frequency sampling & Uniform in wavelength, followed by $f_i=1/\lambda_i$ \\
Forward source & Gaussian, $E_z$; $f_c=1.714286\,c/a$; $f_{\mathrm{width}}=0.571429\,c/a$; cutoff 5; source duration $17.5\,a/c$ \\
Adjoint source & Meep \texttt{FilteredSource}; Nuttall-windowed complex-exponential basis; basis count $=\Nf$ \\
Discretization & 100 pixels/$a$; Courant factor 0.5; $\dtfdtd=0.005\,a/c$ \\
Design region / grid & $5.0a\times0.2a$; material grid $501\times15$ \\
FD DFT workload & Point $E_z$ objective plus active design-region $D_z$ monitor; 11,569 active complex design-region bins per sampled frequency \\
DFT update interval & Meep default automatic decimation (\texttt{decimation\_factor=0}); the temporal update interval is selected internally from the Nyquist requirement of the source/monitor bandwidth \\
Termination & $E_z$ decay at the focus; $10\,a/c$ decay-check window; tolerance $10^{-4}$; common minimum run time $67.5\,a/c$; maximum cap $2000\,a/c$ \\
Execution & AMD Ryzen 9 9950X; one process/rank; one numerical-library thread; process pinned to logical CPU 8 \\
Statistics & Five repetitions per condition in randomized execution order; minimum \texttt{Simulation.run} wall time used for the plotted benchmark \\
\bottomrule
\end{tabularx}
\end{table}

\subsection{Frequency-domain adjoint benchmark}

\subsubsection{Synthesized adjoint-source duration}

The FD benchmark uses Meep's independent-frequency adjoint construction
following the filtered-source formulation of Hammond et al.~\cite{hammond2022high}.
The forward solve is driven by the same broadband Gaussian pulse for every value
of $\Nf$. The adjoint excitation is synthesized using Meep's
\texttt{FilteredSource}, whose basis contains one complex-exponential component
per sampled frequency. For the frequency grid used here, the temporal support
of the filtered source is
\begin{equation}
\TFS = \max_i\left|\frac{1}{f_{i+1}-f_i}\right|
     = \frac{1}{\DFS},
\qquad
\DFS \equiv \min_i |f_{i+1}-f_i|.
\label{eq:tfs}
\end{equation}
Because the frequencies are obtained from a grid that is uniform in wavelength,
$\DFS$ decreases as the wavelength sampling is made denser. Consequently,
$\TFS$ increases strongly with $\Nf$, from $1.75\,a/c$ for $\Nf=2$ to
$486.85\,a/c$ for $\Nf=200$ (Table~\ref{tab:window}).

\begin{table}[htbp]
\caption{Minimum frequency spacing and synthesized \texttt{FilteredSource}
duration used in the FD-adjoint sweep.}
\label{tab:window}
\centering
\small
\begin{tabular}{rrrr}
\toprule
$\Nf$ & $\DFS$ ($c/a$) & Basis count & $\TFS$ ($a/c$) \\
\midrule
2   & 0.5714285714 & 2   & 1.75 \\
5   & 0.1098901099 & 5   & 9.10 \\
10  & 0.0468384075 & 10  & 21.35 \\
15  & 0.0297619048 & 15  & 33.60 \\
20  & 0.0218102508 & 20  & 45.85 \\
30  & 0.0142146411 & 30  & 70.35 \\
50  & 0.0083787181 & 50  & 119.35 \\
100 & 0.0041347943 & 100 & 241.85 \\
150 & 0.0027446137 & 150 & 364.35 \\
200 & 0.0020540207 & 200 & 486.85 \\
\bottomrule
\end{tabular}
\end{table}

The adjoint termination uses the same focal-field decay rule employed in the
forward control, but through Meep's \texttt{until\_after\_sources} pathway.
Therefore, the simulation cannot terminate before the synthesized adjoint
source has ended. At low $\Nf$, the common minimum-run-time floor dominates and
the adjoint simulation remains on a step-count plateau. Once $\TFS$ exceeds
that floor, the simulated time increases approximately as the source duration
plus the final field-decay interval.

\subsubsection{Runtime decomposition into FDTD steps and cost per step}

For each run, the actual Meep simulation time $T_{\mathrm{sim}}$, FDTD time step
$\dtfdtd$, and wall-clock runtime $T_{\mathrm{wall}}$ were recorded. The number
of FDTD updates and the average wall time per update were evaluated as
\begin{equation}
N_{\mathrm{step}} =
\operatorname{round}\!\left(\frac{T_{\mathrm{sim}}}{\dtfdtd}\right),
\qquad
\tstep = \frac{T_{\mathrm{wall}}}{N_{\mathrm{step}}}.
\label{eq:decomp}
\end{equation}
Representative values are listed in Table~\ref{tab:decomp}. As $\Nf$ increases
from 2 to 200, the adjoint simulation grows from 14,007 to 100,050 FDTD updates,
a factor of 7.14. Over the same range, the average cost per update increases
from 0.633 to 8.702 ms/step, a factor of 13.76, and the measured adjoint runtime
increases from 8.861 to 870.634 s, a factor of 98.26. Thus, the runtime growth
contains two distinct contributions associated with the synthesized adjoint
source: its increasing temporal support lengthens the required simulation
window, while the $\Nf$-basis source is also associated with a substantial
increase in per-step computational cost. The latter observation is consistent
with the increasing cost of evaluating the multi-basis filtered source during
the FDTD update; it is not used here to claim a source-level profiling result.

\begin{table}[htbp]
\caption{Representative FD-adjoint runtime decomposition. Each entry is the
minimum wall-clock value among five randomized single-thread runs for that
condition.}
\label{tab:decomp}
\centering
\small
\begin{tabular}{rrrrrr}
\toprule
$\Nf$ & Source end & Simulation time & FDTD steps & Runtime & Runtime / step \\
      & ($a/c$)    & ($a/c$)         &            & (s)     & (ms) \\
\midrule
2   & 1.75   & 70.035  & 14,007  & 8.861   & 0.633 \\
20  & 45.85  & 70.035  & 14,007  & 14.774  & 1.055 \\
50  & 119.35 & 130.065 & 26,013  & 57.170  & 2.198 \\
100 & 241.85 & 250.125 & 50,025  & 210.017 & 4.198 \\
200 & 486.85 & 500.250 & 100,050 & 870.634 & 8.702 \\
\bottomrule
\end{tabular}
\end{table}

\subsubsection{Forward DFT-monitor control}

A separate forward control was used to test whether the observed scaling could
instead be explained by running-DFT accumulation or the associated memory
bandwidth. Two otherwise identical forward simulations were compared: the
standard FD forward solve, which contains the focal-point and design-region DFT
monitors, and a bare-propagation control in which all DFT monitors, incident
normalization, and objective evaluation were removed. The material grid,
geometry, Gaussian source, CPU affinity, FDTD discretization, and focal-field
termination rule were unchanged. The frequency vector in the no-DFT run is
therefore only a batch label and does not modify the field propagation.

At $\Nf=200$, both forward calculations terminate after exactly 14,007 FDTD
updates. The DFT-enabled calculation requires 9.715 s, whereas the no-DFT
control requires 9.025 s (Table~\ref{tab:dftcontrol}). The measured difference
is approximately 0.69 s, or 7.7\% of the bare-propagation runtime. Across the
full sweep, the standard FD-forward runtime changes only weakly with $\Nf$,
whereas the adjoint runtime increases by nearly two orders of magnitude. This
forward--adjoint asymmetry shows that running-DFT accumulation is a measurable
but comparatively small contribution in this benchmark and cannot by itself
explain the observed FD-adjoint scaling.

Importantly, the benchmark uses Meep's default automatic DFT decimation,
with the decimation factor set to 0. Therefore, the result characterizes the
running-DFT cost under the actual Meep configuration used for the paper; it
should not be interpreted as a benchmark in which every DFT accumulator is
explicitly updated at every FDTD time step.

\begin{table}[htbp]
\caption{Forward DFT-monitor control at $\Nf=200$. The two rows use statistically
matched minima from five randomized runs.}
\label{tab:dftcontrol}
\centering
\small
\begin{tabular}{lrrr}
\toprule
Control & FDTD steps & Runtime (s) & Runtime / step (ms) \\
\midrule
Forward with DFT & 14,007 & 9.715 & 0.694 \\
Bare propagation, no DFT & 14,007 & 9.025 & 0.644 \\
Difference & 0 & $\approx 0.69$ & $\approx 0.050$ \\
\bottomrule
\end{tabular}
\end{table}

\subsection{Time-domain adjoint benchmark}

The time-domain comparison uses the same cell, PML, material grid, initial
design, broadband Gaussian excitation, spatial resolution, Courant factor, and
focal-point location as the FD benchmark. Unlike the FD filtered-source
construction, the TD adjoint does not prescribe a separate set of
frequency-domain adjoint amplitudes and therefore its FDTD simulation length
does not increase with $\Nf$. The TD runtime is measured once for a given
sampling interval and is repeated across the $\Nf$ axis only to permit direct
visual comparison with the frequency-sampled FD data.

The forward TD objective used for the runtime benchmark is the temporally
integrated focal-field energy,
\begin{equation}
F_{\mathrm{TD}}
= \frac{1}{2}\sum_n |E_z(t_n)|^2\,\dtfdtd,
\end{equation}
and the time-reversed adjoint source is constructed from the conjugated focal
field history. The adjoint simulation is run to the same total Meep time
reached by the forward calculation. This avoids appending an additional
after-source interval to a time-reversed adjoint source that already spans the
forward simulation history.

Two field-history sampling conditions are benchmarked. In the conventional TD
case, the forward design-region field is sampled at every FDTD time step
($M=1$). In the Nyquist-sampled case, only every 24th forward field sample is
stored and used for gradient accumulation ($M=24$). If $\dot{E}_z$ denotes the
forward-field time derivative entering the discrete adjoint-gradient
correlation, the sampled sequence is
\begin{equation}
\dot{E}_z^{(M)}(m)
= \dot{E}_z(t_{mM}),
\qquad M=1\ \text{or}\ 24.
\end{equation}
The adjoint field is accumulated on the same sampling cadence. The $M=24$
condition is the Nyquist-sampled implementation used for the runtime comparison
reported in the main manuscript. The TD solve therefore preserves the fixed
forward-plus-adjoint simulation horizon while reducing the number of
forward-field samples and gradient-correlation operations.

\subsection{Combined runtime mechanism check}

The combined diagnostics identify two dominant contributions to the
frequency-domain adjoint slowdown as the spectral sampling is refined.
Increasing $\Nf$ reduces the minimum frequency spacing $\DFS$, which increases
the temporal support of the synthesized \texttt{FilteredSource} as
$\TFS=1/\DFS$. Once this source duration exceeds the common minimum-run-time
floor, the number of required FDTD updates increases accordingly. In addition,
the measured wall time per adjoint update also increases with $\Nf$, consistent
with the cost of evaluating the multi-basis synthesized source. By contrast,
the forward DFT-control results show that running-DFT accumulation contributes
only a comparatively small overhead in this benchmark. Thus, for the Meep
implementation tested here, the observed adjoint-runtime scaling is explained
by both the longer temporal simulation horizon and the $\Nf$-dependent
per-step cost associated with the synthesized source.

This interpretation is deliberately implementation-scoped. Alternative
frequency-domain or hybrid adjoint formulations may avoid this particular
filtered-source construction, use different source bases, or employ different
DFT-update strategies. The measured scaling should therefore be interpreted as
a reproducible characterization of the implementation tested in this work,
not as a fundamental complexity bound for frequency-domain adjoint methods.

\bibliographystyle{apsrev4-2}
\bibliography{reference}